\newif\ifdraft
\draftfalse
\ifdraft
\documentclass[manuscript]{aastex631}
\else
\documentclass[twocolumn]{aastex631}
\fi

\let\tablenum\relax

\usepackage{booktabs}
\usepackage{natbib}
\usepackage{mathtools}

\usepackage{savesym}
\savesymbol{tablenum}
\usepackage{siunitx}
\restoresymbol{SIX}{tablenum}

\usepackage{amsmath,here,afterpage}
\usepackage{longtable}
\usepackage{multirow,comment}

\received{}
\revised{}
\accepted{}
\submitjournal{ApJ}

\tighten
\shorttitle{ALMA [CI] Image of the Circumnuclear Disk of the Milky Way}
\shortauthors{Tanaka et al.}
\tighten

\usepackage{float}

\newcommand\pcc{\ifmmode\mathrm{cm^{-3}}\else{$\mathrm{cm^{-3}}$}\fi}
\newcommand\psc{\mathrm{cm^{-2}}}
\newcommand\kelvin{\ifmmode\mathrm{ K}\else K\fi}

\newcommand\kmps{\ifmmode\mathrm{km\,s^{-1}}\else$\mathrm{km\,s^{-1}}$\fi}

\newcommand\pc{\ifmmode\mathrm{pc}\else{pc}\fi}

\newcommand\Msol{\ifmmode{M_\odot}\else${M_\odot}$\fi}

\newcommand\Msun{\ifmmode{M_\odot}\else${M_\odot}$\fi}

\newcommand\Tkin{\ifmmode{T_{\rm kin}}\else{$T_{\mathrm{kin}}$}\fi}
\newcommand\Tex{\ifmmode{T_{\rm ex}}\else{$T_{\mathrm{ex}}$}\fi}
\newcommand\Tturb{\ifmmode{T_{\rm turb}}\else{$T_{\mathrm{turb}}$}\fi}
\newcommand\Td{\ifmmode{T_{\mathrm{d}}\else{$T_{\mathrm{d}}$}\fi}}
\newcommand\Tv{\ifmmode{T_v}\else{${T_v}$}\fi}

\newcommand\Trot{\ifmmode{T_{\rm rot}}\else{$T_{\rm rot}$}\fi}

\newcommand\nH{\ifmmode{n_{\rm H}}\else${n_{\rm H}}$\fi}
\newcommand\nHH{\ifmmode{n_{\rm H_2}}\else{$n_{\rm H_2}$}\fi}
\newcommand\NHH{\ifmmode{N_{\rm H_2}}\else{$N_{\rm H_2}$}\fi}
\newcommand\NHHavg{\ifmmode{\left\langle N_{\rm H_2}\right\rangle}\else{$\left\langle N_{\rm H_2}\right\rangle$}\fi}
\newcommand\nHHavg{\ifmmode{\left\langle n_{\rm H_2}\right\rangle}\else{$\left\langle n_{\rm H_2}\right\rangle$}\fi}
\newcommand\ncrit{\ifmmode{n_{\rm crit}}\else{$n_{\rm crit}$}\fi}

\newcommand\vlsr{\ifmmode{v_{\rm LSR}}\else${v_{\rm LSR}}$\fi}
\newcommand\vlsrm{\ifmmode{v_{\mathrm{LSR},m}}\else${v_{\mathrm{LSR},m}}$\fi}
\newcommand\Mvt{\ifmmode{M_{\rm VT}}\else${M_{\rm VT}}$\fi}
\newcommand\Mhnc{\ifmmode{M_{\rm HNC}}\else${M_{\rm HNC}}$\fi}
\newcommand\Tmb{\ifmmode{T_{\rm MB}}\else{$T_{\mathrm{MB}}$}\fi}

\newcommand\dv {\ifmmode{\rm d}v\else${\rm d}v$\fi}
\newcommand\Eu{\ifmmode{E_{\rm u}}\else{$E_\mathrm{u}$}\fi}

\newcommand\vcol{\ifmmode{v_{\rm col}}\else{$v_{\rm col}$}\fi}

\newcommand\CO{\ifmmode\mathrm{CO}\else$\mathrm{CO}$\fi}
\newcommand\CN{\ifmmode\mathrm{CN}\else$\mathrm{CN}$\fi}

\newcommand\HOCp{\ifmmode{\rm HOC^+}\else${\mathrm HOC^+}$\fi}
\newcommand\CS{\ifmmode{\rm CS}\else${\mathrm CS}$\fi}
\newcommand\SiO{\ifmmode{\rm SiO}\else${\mathrm SiO}$\fi}

\newcommand\HHHp{\ifmmode{\rm H_3^+}\else{${\rm H_3^+}$}\fi}

\newcommand\HCN{\ifmmode{\rm HCN}\else{HCN}\fi}
\newcommand\HCOp{\ifmmode{\rm HCO^+}\else{$\mathrm{HCO^+}$}\fi}
\newcommand\HCNt{\ifmmode{\rm H{^{13}C}N}\else{$\mathrm{H{^{13}C}N}$}\fi}
\newcommand\HNC{\ifmmode{\rm HNC}\else{$\mathrm{HNC}$}\fi}
\newcommand\HCCCN{\ifmmode{\rm HC_3N}\else{$\mathrm{HC_3N}$}\fi}
\newcommand\HCOpt{\ifmmode{\rm H^{13}CO^+}\else{$\mathrm{H^{13}CO^+}$}\fi}

\newcommand\Cn{\ifmmode {\rm C^0}\else $\mathrm{C^0}$\fi}

\newcommand\COt{\ifmmode{\rm {^{13}CO}}\else{$\mathrm{^{13}CO}$}\fi}
\newcommand\Ct{\ifmmode{\rm {^{13}C}}\else{$\mathrm{^{13}C}$}\fi}
\newcommand\Ctw{\ifmmode{\rm {^{12}C}}\else{$\mathrm{^{12}C}$}\fi}
\newcommand\ammonia{\ifmmode{\rm NH_3}\else{$\rm NH_3$}}

\newcommand\NNHp{\ifmmode{\rm N_2H^+}\else{$\mathrm{N_2H^+}$}\fi}
\newcommand\HHCS{\ifmmode{\rm H_2CS}\else{$\mathrm{H_2CS}$}\fi}
\newcommand\CtS{\ifmmode{\rm ^{13}CS}\else{$\mathrm{^{13}CS}$}\fi}
\newcommand\OCS{\ifmmode{\rm OCS}\else{$\mathrm{OCS}$}\fi}
\newcommand\methanol{\ifmmode{\rm CH_3OH}\else{$\mathrm{CH_3OH}$}\fi}
\newcommand\pformaldehyde{\ifmmode{p\mbox{-}\rm H_2CO}\else{$p$-$\mathrm{H_2CO}$}\fi}
\newcommand\formaldehyde{\ifmmode{\rm H_2CO}\else{$\mathrm{H_2CO}$}\fi}
\newcommand\pfx{\ifmmode{3_{21}--2_{20}}\else{$3_{21}--2_{20}$}\fi}
\newcommand\pfy{\ifmmode{3_{22}--2_{21}}\else{$3_{22}--2_{21}$}\fi}
\newcommand\HHO{\ifmmode{\rm H_2O}\else{$\mathrm{H_2O}$}\fi}

\newcommand\JJ[2]{\ifmmode{\mbox{{\it J}={#1}\mbox{--}{#2}}}\else{{\it J}={#1}--{#2}}\fi}
\newcommand\JJx[2]{\ifmmode{\mbox{{#1}\mbox{--}{#2}}}\else{{#1}--{#2}}\fi}
\newcommand{\NN}[1]{\ifmmode N\left(#1\right)\else$N\left(#1\right)$\fi}
\newcommand\JK[4]{\ifmmode{{J_K}=#1_{#2}\mbox{--}#3_{#4}}\else{${\it J_K}=#1_{#2}\mbox{--}#3_{#4}$}\fi}

\newcommand\CIa{\ifmmode{^3}P_1\mbox{--}{^3}P_0\else${^3}P_1\mbox{--}{^3}P_0$\fi}
\newcommand\CIb{\ifmmode{^3}P_2\mbox{--}{^3}P_1\else${^3}P_2\mbox{--}{^3}P_1$\fi}

\newcommand\gl{\ifmmode l\else{\it l}\fi}
\newcommand\gb{\ifmmode b\else{\it b}\fi}

\newcommand\sgras{$\mathrm{Sgr A^{*}}$}

\newcommand\CLb{CO$-0.30$$-0.07$}
\newcommand\theObj\CLb

\newcommand\Dv{\ifmmode{\Delta v}\else{$\Delta v$}\fi}
\newcommand\Dvheat{\ifmmode{\Delta v_{\rm heat}}\else{$\Delta v_{\rm heat}$}\fi}
\newcommand\Sv{\ifmmode{\sigma v}\else{$\sigma v$}\fi}
\newcommand\vc{\ifmmode{\left<v\right>}\else{$\left<v\right>$}\fi}

\newcommand{\PV}{{\it P}--{\it V}}
\newcommand{\avir}{\ifmmode{\alpha_{\rm vir}}\else{$\alpha_{\rm vir}$}\fi}
\newcommand{\ntild}{\ifmmode{n^*}\else{$n^*$}\fi}

\newcommand\Sdust{\ifmmode{S_{500}}\else{$S_{500}$}\fi}
\newcommand\II[1]{\ifmmode{I_{#1}}\else{$I_{#1}$}\fi}

\newcommand\myvector[1]{\ifmmode{\mbox{\boldmath ${#1}$}}\else{\boldmath {${#1}$}}\fi}
\newcommand{\Rt}{\ifmmode{{R_{13}}}\else${R_{13}}$\fi}
\newcommand{\Iobs}{\ifmmode{I}\else${I}$\fi}
\newcommand{\Icalc}{\ifmmode{{F}\left(\pv\right)}\else${{F}\left(\pv\right)}$\fi}
\newcommand{\Icalci}{\ifmmode{{F}\left(\pv_{i}\right)}\else${{F}\left(\pv_{i}\right)}$\fi}
\newcommand{\Icalcv}{\ifmmode{\myvector{F}\left(\pv\right)}\else${\myvector{F}\left(\pv\right)}$\fi}
\newcommand{\xmol}{\ifmmode{{x_{\rm mol}}}\else${x_{\rm mol}}$\fi}
\newcommand{\xmolp}[1]{\ifmmode{{x_{\rm mol}\left(#1\right)}}\else${x_{\rm mol}\left(\mbox{#1}\right)}$\fi}
\newcommand{\ff}{\ifmmode{\Phi}\else${\Phi}$\fi}
\newcommand{\fff}{\ifmmode{\phi}\else${\phi}$\fi}
\newcommand{\fcal}{\ifmmode{f_{\rm cal}}\else${f_{\rm cal}}$\fi}

\newcommand{\Ea}{\ifmmode\epsilon^{\rm a}\else$\epsilon_{\rm a}$\fi}
\newcommand{\Em}{\ifmmode\epsilon\else$\epsilon$\fi}

\newcommand{\sa}{\ifmmode\sigma\else$\sigma$\fi}
\newcommand{\sm}{\ifmmode\sigma\else$\sigma$\fi}
\newcommand{\scal}{\ifmmode\sigma_{\rm cal}\else$\sigma_{\rm cal}$\fi}

\newcommand{\av}{\myvector{a}}

\newcommand{\pv}{\myvector{p}}

\newcommand{\PDF}[1]{{\ifmmode P(#1) \else $P(#1)$ \fi}}

\newcommand\RRx{\ifmmode R_{43}\else$R_{43}$\fi}
\newcommand\IIx{\ifmmode I_{13}\else$I_{13}$\fi}
\newcommand\Ihnc{\ifmmode I_{\HNC}\else$I_{\HCN}$\fi}
\newcommand\Ihcn{\ifmmode I_{\HCN 43}\else$I_{\HCN 43}$\fi}
\newcommand\Ihcnt{\ifmmode I_{\HCNt}\else$I_{\HCNt}$\fi}
\newcommand\Ihcccn{\ifmmode I_{\HCCCN}\else$I_{\HCCCN}$\fi}

\newcommand\Lhnc{\ifmmode L_{\HNC}\else$L_{\HCN}$\fi}
\newcommand\Lhcn{\ifmmode L_{\HCN 43}\else$L_{\HCN 43}$\fi}
\newcommand\Lhcnt{\ifmmode L_{\HCNt}\else$L_{\HCNt}$\fi}
\newcommand\Lhcccn{\ifmmode L_{\HCCCN}\else$L_{\HCCCN}$\fi}

\newcommand\dNdv{\ifmmode {\mathrm{d}N}/{\mathrm{d}v}\else${\mathrm{d}N}/{\mathrm{d}v}$\fi}
\newcommand\dNHdv{\ifmmode \frac{\mathrm{d}N_{\rm H_2}}{\mathrm{d}v}\else$\frac{\mathrm{d}N_{\rm H_2}}{\mathrm{d}v}$\fi}
\newcommand\Xdvdr{\ifmmode {{X}/{\frac{\mathrm{d}v}{\mathrm{d}r}}}\else{${X}/\frac{\mathrm{d}v}{\mathrm{d}r}$}\fi}

\newcommand\Nobs{\ifmmode{N_{\rm obs}}\else{$N_{\rm obs}$}\fi}
\newcommand\Nmol{\ifmmode{N_{\rm mol}}\else{$N_{\rm mol}$}\fi}
\newcommand\Np{\ifmmode{N_{p}}\else{$N_{p}$}\fi}
\newcommand\Nl{\ifmmode{N_{l}}\else{$N_{l}$}\fi}

\newcommand\dvdr{\ifmmode{{\mathrm d}v/{\mathrm d}r}\else{${\mathrm d}v/{\mathrm d}r$}\fi}

\newcommand\Mmag{\ifmmode M_\Phi\else$M_\Phi$\fi}
\newcommand\nth{\ifmmode {n_{\mathrm{th}}}\else$n_{\mathrm{th}}$\fi}
\newcommand\SFRff{\ifmmode {\mathrm{SFR_{ff}}}\else$\mathrm{SFR_{ff}}$\fi}

\newcommand\zCR{\ifmmode \zeta_{\mathrm{CR}}\else$\zeta_{\mathrm{CR}}$\fi}
\newcommand\xe{\ifmmode x_{\mathrm{e}}\else$x_{\mathrm{e}}$\fi}
\newcommand\Qc{\ifmmode Q_\mathrm{c} \else $Q_\mathrm{c}$\fi}
\newcommand\Qnc{\ifmmode Q_\mathrm{nc} \else $Q_\mathrm{nc}$\fi}

\newcommand\logt{\ifmmode \log_{10}\else $\log_{10}$\fi}
\newcommand\parsec{\ifmmode pc\else $\mathrm{pc}$\fi}
\newcommand\Prob[1]{\ifmmode \mathrm{Pr}\left(#1\right) \else $\mathrm{Pr}\left(#1\right)$ \fi}
\newcommand\avpv{\ifmmode \av\cdot\pv \else $\av\cdot\pv$\fi}

\newcommand\fsf{\ifmmode {f_{\mathrm{SF}}}\else${f_{\mathrm{SF}}}$\fi}
\newcommand\tff{\ifmmode {t_{\mathrm{ff}}}\else${t_{\mathrm{ff}}}$\fi}
\newcommand\epsff{\ifmmode {\epsilon^*_{\mathrm{ff}}}\else${\epsilon^*_{\mathrm{ff}}}$\fi}
\newcommand\etasf{\ifmmode {\eta_{\mathrm{SF}}}\else${\eta_{\mathrm{SF}}}$\fi}
\newcommand\etadense{\ifmmode {\eta_{\mathrm{dense}}}\else${\eta_{\mathrm{dense}}}$\fi}
\newcommand\ith{\ifmmode {{i}^{\mathrm{th}}}\else ${{i}^{\mathrm{th}}}$\fi}
\newcommand\Npar{\ifmmode {N_{\rm param}}\else ${N_{\mathrm{param}}}$\fi}

\newcommand\SigmaH{\ifmmode {\Sigma_{\rm H_2}}\else ${\Sigma_{\mathrm{H_2}}}$\fi}

\newcommand\Myr{\ifmmode {\rm Myr}\else Myr\fi}
\newcommand\rmaj{\ifmmode{r_{\mathrm{maj}}}\else${r_{\mathrm{maj}}}$\fi}
\newcommand\rmin{\ifmmode{r_{\mathrm{min}}}\else${r_{\mathrm{min}}}$\fi}

\newcommand\PPV{\textit{P}--\textit{P}--\textit{V}}
\newcommand\PP{\textit{P}--\textit{P}}

\newcommand\ttr{\ifmmode{\tau_{\mathrm{cl}}}\else $\tau_{\mathrm{cl}}$\fi}

\newcommand\CI{[\ion{C}{1}]}

\newcommand\Tci{\ifmmode T({\rm CI})\else $T({\rm CI})$\fi}
\newcommand\Tct{\ifmmode T(\COt)\else $T({\COt})$\fi}

\newcommand\dCI{\ifmmode {\mathrm{d}T}\else {$\mathrm{d}T$} \fi}
\newcommand\dCIx{\ifmmode {\mathrm{d}T_{1\text{--}0}}\else {$\mathrm{d}T_{1\text{--}0}$} \fi}
\newcommand\dCIy{\ifmmode {\mathrm{d}T_{2\text{--}1}}\else {$\mathrm{d}T_{2\text{--}1}$} \fi}
\newcommand\RCI{\ifmmode R\else $R$\fi}

\newcommand\Rco{\ifmmode R_{\rm 2\mathchar`-1/1\mathchar`-0}\else $R_{\rm 2\mathchar`-1/1\mathchar`-0}$\fi}

\newcommand\molH{\ifmmode\mathrm{H_2}\else$\mathrm{H_2}$\fi}
\newcommand\ps{\ifmmode\mathrm{s^{-1}}\else$\mathrm{yr^{-1}}$\fi}
\newcommand\yr{\ifmmode\mathrm{yr}\else$\mathrm{yr}$\fi}

\newcommand{\jkpb}{\ifmmode \mathrm{Jy\,beam^{-1}}\else$\mathrm{Jy\,beam^{-1}}$\fi}

\newcommand{\Sdense}{\ifmmode S_\mathrm{dense}\else $S_\mathrm{dense}$\fi}
\newcommand{\Snorm}{\ifmmode S_\mathrm{norm}\else $S_\mathrm{norm}$\fi}
\newcommand{\SCS}{\ifmmode \tilde{S}\left(\mathrm{CS}\right)\else $\tilde{S}\left(\mathrm{CS}\right)$\fi}
\newcommand{\SCI}{\ifmmode \tilde{S}\left(\mathrm{CI}\right)\else $\tilde{S}\left(\mathrm{CI}\right)$\fi}

\newcommand{\Vrad}{\ifmmode V_\mathrm{rad}\else $V_\mathrm{rad}$\fi}
\newcommand{\Vradm}{\ifmmode V_{\mathrm{rad},m}\else $V_{\mathrm{rad},m}$\fi}
\newcommand{\Vradn}{\ifmmode V_{\mathrm{rad},n}\else $V_{\mathrm{rad},n}$\fi}
\newcommand{\Vrot}{\ifmmode V_\mathrm{rot}\else $V_\mathrm{rot}$\fi}
\newcommand{\Vmax}{\ifmmode V_\mathrm{max}\else $V_\mathrm{max}$\fi}
\newcommand{\Rmin}{\ifmmode R_\mathrm{min}\else $R_\mathrm{min}$\fi}
\newcommand{\Vphi}{\ifmmode V_\phi\else $V_\phi$\fi}

\newcommand{\Rproj}{\ifmmode R_\mathrm{proj}\else $R_\mathrm{proj}$\fi}
\newcommand{\PAproj}{\ifmmode \phi_\mathrm{proj}\else $\phi_\mathrm{proj}$\fi}
\newcommand{\mstar}{\ifmmode{m_\mathrm{1pc}}\else $m_\mathrm{1pc}$\fi}
\newcommand{\mbh}  {\ifmmode{m_\mathrm{BH}}\else $m_\mathrm{BH}$\fi}

\newcommand{\alpIRCS}{\ifmmode\alpha\else $\alpha$\fi}
\newcommand{\deltIRCS}{\ifmmode\delta\else $\delta$\fi}
\newcommand{\Dci}{\ifmmode D_{\mathrm{CI}}\else $D_{\mathrm{CI}}$\fi}

\newcommand{\Mmol}{{\ifmmode M_{\mathrm{CI+CS}}\else $M_{\mathrm{CI+CS}}$\fi}}
\newcommand{\Lci}{{\ifmmode L_{[\mathrm{CI}]}\else $L_{[\mathrm{CI}]}$\fi}}
\newcommand{\Xci}{{\ifmmode X_{[\mathrm{CI}]}\else $X_{[\mathrm{CI}]}$\fi}}
\newcommand{\Lcs}{{\ifmmode L_{\mathrm{CS7\mathchar`-6}}\else {$L_{\mathrm{CS7\mathchar`-6}}$}\fi}}
\newcommand{\Xcs}{{\ifmmode X_{\mathrm{CS7\mathchar`-6}}\else {$X_{\mathrm{CS7\mathchar`-6}}$}\fi}}

\newcommand{\tightcdot}{\mspace{-2mu}\cdot\mspace{-1mu}}
\newcommand{\Mlow}{{\ifmmode M_{\mathrm{low}}\else $M_{\mathrm{low}}$\fi}}
\newcommand{\Mhigh}{{\ifmmode M_{\mathrm{high}}\else $M_{\mathrm{high}}$\fi}}
\newcommand{\Rlh}{{\ifmmode R_{\mathrm{low}/\mathrm{high}}\else $\Rlh$\fi}}
\newcommand{\lumunit}{{\ifmmode\kelvin\cdot\kmps\tightcdot\pc^2\else$\lumunitTab$\fi} }
\newcommand{\lumunitTab}{{\ifmmode\kelvin\tightcdot\kmps\tightcdot\pc^2\else$\lumunitTab$\fi} }
\newcommand{\Xunit}{{\ifmmode\Msun\cdot\pc^{-2}\left(\kelvin\cdot\kmps\right)^{-1}\else$\Xunit$\fi} }
\newcommand{\XunitTab}{{\ifmmode\frac{\Msun\tightcdot\pc^{-2}}{\kelvin\tightcdot\kmps}\else$\XunitTab$\fi} }
\newcommand{\rclump}{{\ifmmode{r_\mathrm{clump}}\else{$\rclump$}\fi}}
\newcommand{\nroche}{{\ifmmode{n_\mathrm{Roche}}\else{$\nroche$}\fi}}
\newcommand{\ttidal}{{\ifmmode{\tau_\mathrm{TD}}\else{$\ttidal$}\fi}}

\newcommand{\dRA}{{\ifmmode{\mathrm{dRA}}\else{$\dRA$}\fi}}
\newcommand{\dDec}{{\ifmmode{\mathrm{dDec}}\else{$\dDec$}\fi}}
\newcommand{\Tdust}{{\ifmmode{T_\mathrm{dust}}\else{$\Tdust$}\fi}}

\newcommand{\myrevision}[1]{{\textcolor{red}{#1}}}

\renewcommand{\myrevision}[1]{#1}

\newenvironment{revision}{\color{black}}{\color{black}}

\newcommand{\bestAlp}{\ang{27.3;;}}
\newcommand{\bestI}{\ang{72.7;;}}

\begin{document}


\title{ALMA [CI] Image of the Circumnuclear Disk of the Milky Way: Inflowing Low-density Molecular Gas}
\author[0000-0001-8153-1986]{Kunihiko Tanaka}
\email{ktanaka@keio.jp}
\affil{Department of Physics, Faculty of Science and Technology, Keio University, 3-14-1 Hiyoshi, Yokohama, Kanagawa 223-8522 Japan}

\author{Makoto Nagai}
\affil{National Astronomical Observatory of Japan, 2-21-1 Osawa, Mitaka, Tokyo 181-8588, Japan}
\author{Kazuhisa Kamegai}
\affil{Public Relations Center, National Astronomical Observatory of Japan, 2-21-1 Osawa, Mitaka, Tokyo 181-8588, Japan}



\begin{abstract}
We present ALMA [\ion{C}{1}]~$^3P_1$--$^3P_0$ imaging of the central $6.6\times4.2~\mathrm{pc}^2$ region of the Galaxy encompassing the circumnuclear disk (CND).
The data reveal low-density ($n_\mathrm{H_2}\sim10^3~$cm$^{-3}$) molecular gas with inward motion, widespread both inside and outside the CND.
The normalized [\ion{C}{1}] to CS~7--6 intensity difference decreases inwardly from $R=4$~pc to 1.7~pc and azimuthally along the CND's rotation, likely tracing paths of low-density gas inflow.
By projecting spaxels into orbital coordinates assuming a velocity field model, we identify four kinematic features: a pair of spiral outer streamers toward the CND, inner streamers extending to 0.5~pc from Sgr~A$^*$, an outer disk at $ R\sim3$--6~pc, and the rotating ring at $R=2$~pc.
$P$--$P$--$V$ correlation between the inner streamers and H42$\alpha$ indicates gas supply to the mini-spiral through the western arc (WA) and northern arm (NA).
The total inflowing mass is $1.5\times10^4~M_\odot$, 1.7 times \myrevision{the mass of the rotating ring}.
The identified flows can be organized into two main pathways connecting the CND exterior and interior: ``WA flow'' feeding the mini-spiral WA via the CND, and ``NA flow'' bypassing \myrevision{the purely rotating orbit}.
The inflow rate along the former is approximately constant (0.1--0.16~$M_\odot~\mathrm{yr}^{-1}$), implying a CND dwelling time comparable to its orbital period and supporting the CND's transient nature.
We also identify two [\ion{C}{1}]-bright clumps (CBCs) lacking dense-gas counterparts near the contact point between the northern outer streamer and the CND.
Apparently intact against tidal disruption despite subcritical densities, the CBCs may represent a chemically young phase shortly after formation in colliding flows.
\end{abstract}

\section{Introduction}

\begin{revision}
The Milky Way’s circumnuclear disk (CND) is a dense molecular ring rotating at a radius of $\sim$2~pc from the nuclear supermassive black hole \sgras. 
The CND is regarded as a Galactic analog of circumnuclear dense-gas condensations commonly found in galactic nuclei, which are considered reservoirs of material that fuel nuclear activity and circumnuclear star formation.
Although the present Galactic center does not exhibit quasar activity or ongoing massive star formation, observations suggest a past active phase of \sgras\ and a nuclear starburst event that formed the central stellar cluster \citep[e.g.,][]{Paumard2006,Zubovas2011,Heywood2019}.
As the closest dense-gas condensation to \sgras, the CND is crucial for understanding the relationship between these events and the properties of the circumnuclear dense gas, providing the nearest template for quasar and nuclear starburst systems.
\end{revision}
Although the CND appears as an isolated structure in dense-gas tracer images, recent observations increasingly suggest that it is physically connected to ionized and molecular gas both inside and outside the CND ring.
Molecular cloud features, referred to as streamers and anomalies, are distributed from just outside the CND ring to the 20-\kmps\ and 50-\kmps\ clouds \citep{Coil1999,Christopher2005,Montero-Castano2009,Liu2012,Takekawa2017,Hsieh2017,Tsuboi2018}.
Their line-of-sight velocities often deviate from the purely rotational motion exhibited by the CND ring, indicating the presence of radial motion.
At the inner edge of the CND ring lies the mini-spiral, a system of dense ionized gas structures that follow high-eccentricity Keplerian orbits or spiral streams \citep{Zhao2009,Irons2012,Tsuboi2017a}.
The consistent magnetic field structure across the CND and the central cavity inside the CND \citep{Hsieh2018}, along with the spatial and velocity continuity of molecular and hydrogen recombination lines at their interfaces \citep{Lo1983,Genzel1985,Serabyn1985,Guesten1987,Christopher2005,Tsuboi2018}, suggest a physical continuity between the ionized inner surface of the CND and the mini-spiral.
Recent observations have also revealed cloudlets and filaments inside the CND \citep{Yusef-Zadeh2013, Moser2016, Goicoechea2018a, Hsieh2019}, which were previously thought to be dominated by ionized gas, though their kinematic relationship to the CND is not fully established.

\begin{revision}
These molecular and ionized features represent molecular inflow toward the Galactic center.
The CND and its associated streamers or anomalies, as well as the mini-spiral, are thought to be immediate products of one or multiple GMC capture events \citep{Sanders1998,Vollmer2001a,Vollmer2001b,Vollmer2002,Bonnell2008,Wardle2008,Namekata2009,Hobbs2009,Alig2013,Mapelli2016,Ballone2019,Tress2020}, which are also considered to have formed the young stellar disks within the central 0.5~\pc\ (the `\textit{in-situ}'~formation scenario; \citealt{Paumard2006,Bonnell2008,Hobbs2009,Alig2013,Mapelli2016}).
The approximate alignment of the CND’s orbital plane with that of the central molecular zone (CMZ) \citep{Jackson1993,Sofue2025} supports the idea that the CND originated from CMZ clouds, whose infall could be facilitated by dynamical friction, SN feedback, the self-gravity of the CMZ ring, or interaction with the nested bar potential \citep{Stark1991,Sanders1998,Namekata2009,Tress2020}.
The average dense gas in the CND is tidally subcritical \citep{Guesten1987,Requena-Torres2012}, and the present ring structure is considered a transient feature, although opposing arguments also exist \citep{Vollmer2001a,Vollmer2001b,Blank2016}.
The CND transfers mass to the central cavity inside it through cloud-cloud collisions within the CND and between the CND and streamers, viscous processes, and mixing with ejecta from stellar winds and supernovae \citep{Blank2016,Solanki2023,Barna2025}.

\end{revision}

The CND and its associated molecular cloud features, namely, the streamers and anomalies, have been primarily observed using high-critical density (\ncrit) molecular lines such as \ammonia, HCN, CS, and high-$J$ CO \citep{Guesten1987,Jackson1993, Marr1993, Coil1999, Coil2000, Christopher2005, Herrnstein2005, Montero-Castano2009, Liu2012, Martin2012, Liu2013, Tsuboi2018, Hsieh2019, Hsieh2021a}, which target molecular cloud components with high density and high temperature ($\Tkin\gtrsim200~\kelvin$, $\nHH\gtrsim10^4~\pcc$; \citealt{Requena-Torres2012}).
However, these observations were not sufficiently sensitive to the low-excitation molecular gas ($\Tkin\sim20\mbox{--}30~\kelvin$, $\nHH\sim10^3~\pcc$; \citealt{Tanaka2018b,Tanaka2021}) that pervades the entire CND.
Such low-excitation gas in the central $\sim10$~pc region has been observed to extend into areas devoid of high-density tracer emissions, including the apparent gap between the 50-\kmps\ cloud and the CND \citep{Tanaka2011, Garcia2016, Tanaka2021}, underscoring the importance of low-density tracers for studying molecular gas flows associated with the CND.
A cold dust component ($\Tdust\sim23.5$~K), likely corresponding to the low-excitation molecular gas, has also been identified in far-infrared dust SEDs, with a mass approximately twice that of the warmer component ($\Tdust\sim~44.5$~K; \citealt{Etxaluze2011}).
Despite its significance in mass and kinematics, the low-excitation gas remains less understood than the high-excitation component, primarily due to the difficulty of low-$J$ CO and CO isotopologue observations \citep[e.g.,][]{Liu2012}. These lines are severely contaminated by foreground and background emission from non-CND gas in the CMZ and spiral arms, in particular at low \vlsr\ ($\left|\vlsr\right|\lesssim100~\kmps$). 

This paper presents interferometric imaging of the \CI~\CIa\ transition of atomic carbon (\Cn)\footnote{\myrevision{This paper uses the terminology where \Cn\ refers to the atomic species itself, while \CI\ denotes its transitions.}} to investigate low-density molecular gas associated with the CND.
\Cn\ is one of the most abundant atomic/molecular species observed in interstellar molecular gas, whose typical abundance ratio to CO is $\sim 0.1$ in the Galactic disk region and $\lesssim 0.5$ in the CMZ (\citealt{Tanaka2011,Tanaka2021}; and referenced therein). 
Single-dish \CI\ observations \citep{Tanaka2011, Garcia2016, Tanaka2021} have revealed a highly elevated $N(\Cn)$/$N(\CO)$ abundance ratio of $\sim2$ in the CND, presumably due to a high cosmic-ray dissociation rate and/or the effect of time-dependent chemistry. 
In addition to the rich \Cn\ abundance, the low \ncrit\ ($\ncrit=10^3~\pcc$) and low upper-state energy ($\Eu/k_\mathrm{B}= 23.6~\kelvin$) of the \CI~\CIa\ transition make it an ideal tracer of low-density, low-temperature gas, while being relatively unaffected by contamination from clouds outside the CND. 
We conducted interferometric \CI~\CIa\ imaging observations over a $6.5\times3.5~\pc^2$ elliptical area encompassing the CND and a major portion of the associated streamers, with the aim of delineating a more complete picture of the CND, CND-associated features, and their relation to the mini-spiral.

The remainder of this paper is structured as follows. Section~\ref{section:observation} describes the observations conducted with the Atacama Large Millimeter/submillimeter Array (ALMA) and the data reduction. Section~\ref{section:results} presents the \CI\ image along with the byproduct CS~10--9 image, highlighting the discovery of several \CI-bright features. This section also includes an analysis of the three-dimensional kinematics, identification of streamer components, measurements of the spatial variation in the \CI-to-CS ratio, and mass estimates. Section~\ref{section:discussion} discusses the global molecular gas flows connecting the CND and the central cavity, based on these analyses. Finally, Section~\ref{section:summary} summarizes the main results of this study. Throughout this paper, we assume a distance to \sgras\ of $D=8.18$~kpc \citep{GravityCollaboration2019}. The coordinate origin of all images is taken at \sgras, ($\mathrm{17^h45^m40^s.04}$, $\ang{-29;00;28.17}$).
\myrevision{
The main products of the analysis presented in this paper are available at our GitHub repository: \url{https://github.com/kutanaka/Tanaka-et-al.-2026-ApJ-online-data}.
}

\section{Observation and Data Reduction\label{section:observation}}
\begin{deluxetable}{lcc}
\tabletypesize{\footnotesize}
\tablecolumns{3}
\tablehead{
\colhead{\shortstack{Date\\{}}} & \colhead{\shortstack{Mosaic Field\\{}}} & \colhead{\shortstack{{}\\On-source Integration\\ Time (s)}}   
}
\tablecaption{ALMA Execution Blocks \label{table:EBs}}
\startdata 
15-Jun-2016 & 7m-S   & 1508 \\
18-Jul-2016 & 7m-S   & 1508 \\
22-Jul-2016 & 7m-S   & 1568  \\
29-Jul-2016 & 12m-S  & 1913 \\
12-Aug-2016 & 7m-N & 1467  \\
16-Aug-2016 & 7m-N  & 1467 \\
17-Aug-2016 & 7m-N  & 1467 \\
18-Aug-2016 & 12m-N  & 1954 
\enddata
\end{deluxetable}

\begin{deluxetable}{lcccc}
\tabletypesize{\footnotesize}
\tablecolumns{5}
\tablecaption{Observed Frequencies \label{table:spws}}
\tablehead{
\colhead{} &\colhead{\shortstack{\\{}Center\\ (GHz)}} & \colhead{\shortstack{Bandwidth\\ (GHz)}} & \colhead{\shortstack{Channel bin\\ (GHz)}} & \colhead{\shortstack{Transitions\\{}}}
} 
\startdata 
SPW0 & 492.16065 & 1.98 & \phn1.13 & \CI~$\CIa$ \\
SPW1 & 490.20000 & 1.98 & 31.25 & $\CS~\JJ{10}{9}$   \\
SPW2 & 478.20000 & 1.98 & 31.25 &    \\
SPW3 & 480.02000 & 1.98 & 31.25 &
\enddata
\end{deluxetable}

\begin{figure}
    \centering
    \epsscale{1.1}
    \plotone{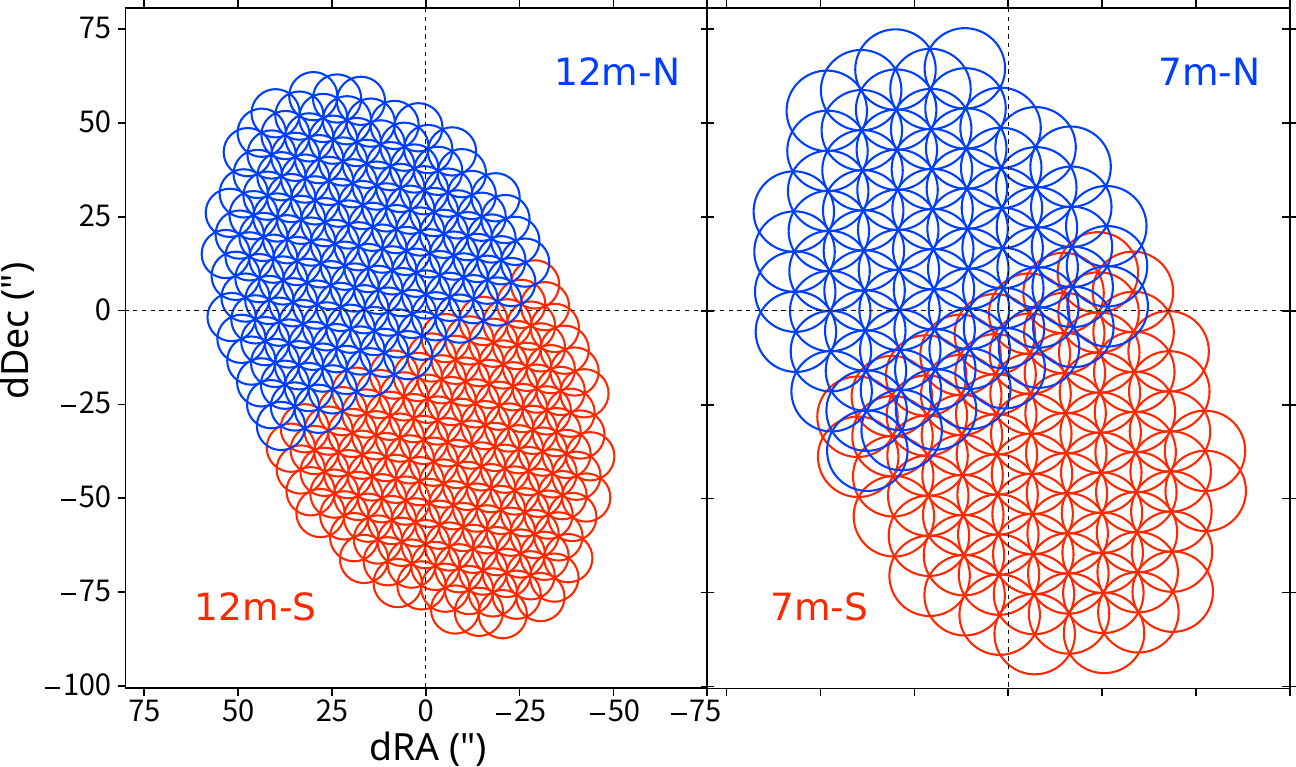}
    \caption{ALMA fields of view (FoVs) of the 12-m array (left) and 7-m array (right) observations.
    The circles indicate the primary beams at 490 GHz (\ang{;;13} and \ang{;;21}\ FWHMs for the 12-m and 7-m arrays, respectively). The northern (12m-N and 7m-N) and southern fields (12m-S and 7m-S) were observed in separate scheduling blocks. } 
    \label{fig:fov}
\end{figure}

\begin{figure}
    \centering
    \epsscale{1.1}
    \plotone{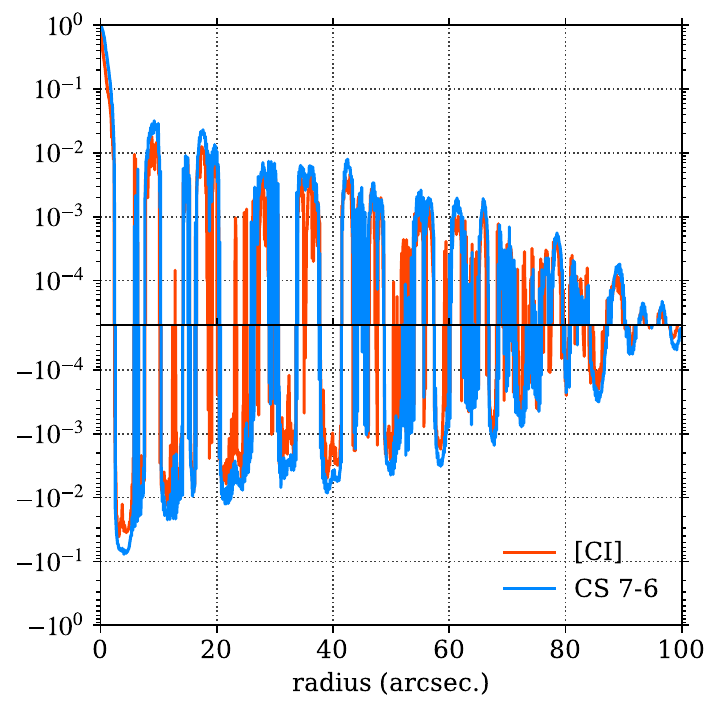}
    \caption{Azimuthally averaged profiles of the point spread functions (PSFs) of the \CI\ and CS~10--9 data.}
    \label{fig:psf}
\end{figure}

The ALMA observations were conducted in 2015 (project code: 2015.1.01040.S; PI: Kunihiko Tanaka).
An approximately elliptical region shown in Figure \ref{fig:fov} was covered by mosaics with 285 and 144 field of views (FoVs) of the 12-m and 7-m array observations, respectively.   
The observation was performed with 8 execution blocks (EBs) listed in Table \ref{table:EBs}. 
Four spectral windows (SPWs) listed in Table \ref{table:spws} were simultaneously observed.
The main target line \CI~\CIa\ (492.1706 GHz) was observed in SPW0 with a 1.13 MHz channel separation, corresponding to a $\sim 0.7~\kmps$ velocity channel width.  The other SPWs were configured for continuum observations, which included the CS~\JJ{10}{9} (489.7509 GHz) transition as a bi-product.

The data reduction was performed by using the Common Astronomy Software Applications \myrevision{(CASA; \citealt{CASA2022})} package developed by the National Radio Astronomy Observatory.  Flagging low-quality data, correction for atmospheric conditions, and calibration of the visibility data were done \myrevision{via CASA pipeline using the reduction script included in the ALMA dataset, ``scriptForPI.py''}.
Neptune and J1751+0939 were used as flux calibrators, and J1924-2914 and J1733-3722 were used as the bandpass and complex gain calibrators, respectively.
Phase-only self-calibration was performed using the pseudo-continuum image integrated over the entire bandwidth.
The continuum and line components were separated using the the {\tt uvcontsub} task of CASA. 

The position--position--velocity (\PPV) cube was created from the continuum-subtracted SPW0 data using the {\tt tclean} task.
The Briggs weighting with the robustness parameter of 2 was chosen as the weighting function.  
Although the antenna configuration of the 12-m array allowed the maximum spatial resolution of $\sim\ang{;;0.3}$, we applied a $uv$-taper of $0.8''$ to obtain a sufficient signal-to-noise ratio. 
The resultant synthesized beam size of the combined image from the 12-m and 7-m array data is $0.86''\times0.64''$ with the position angle (PA) of $\ang{-78.4}$.  \myrevision{The azimuthally averaged profile of the point spread function (PSF) is presented in Figure \ref{fig:psf}.} 
The velocity coverage and resolution are $\pm 250~\kmps$ \vlsr\ and $2~\kmps$, respectively.  
The rms noise level \myrevision{per 2-\kmps\ velocity channel} after primary beam correction is 0.14 \jkpb\ in the southern part of the observation field and 0.10 \jkpb\ in the northern part.
The maximum recoverable angular scale is \ang{;;13.6}.   
The missing flux measured using the single-dish observations with the ASTE 10-m telescope \citep{Tanaka2021} is 43\%. 

The CS~\JJx{10}{9}\ cube was created from the continuum-subtracted SPW1 data for the same parameters as the \CI\ cube but with 20~\kmps\ velocity channel separation and a \ang{;;2} $uv$-taper.
The synthesized beam size is $\ang{;;1.95}\times\ang{;;1.46}$ with a $-57.4^\circ$ PA.
\myrevision{The PSF of the CS data is shown in Figure \ref{fig:psf}.} 
The rms noise levels \myrevision{per 20-\kmps\ velocity channel} after the primary beam corrections are 0.094 \jkpb\ and 0.070 \jkpb in the southern and northern fields.

\section{results and analysis\label{section:results}}

\subsection{\CI-bright features \label{subsection:results:images}}

\begin{figure*}[ttt]
  \begin{center}
\epsscale{1.}
    \plotone{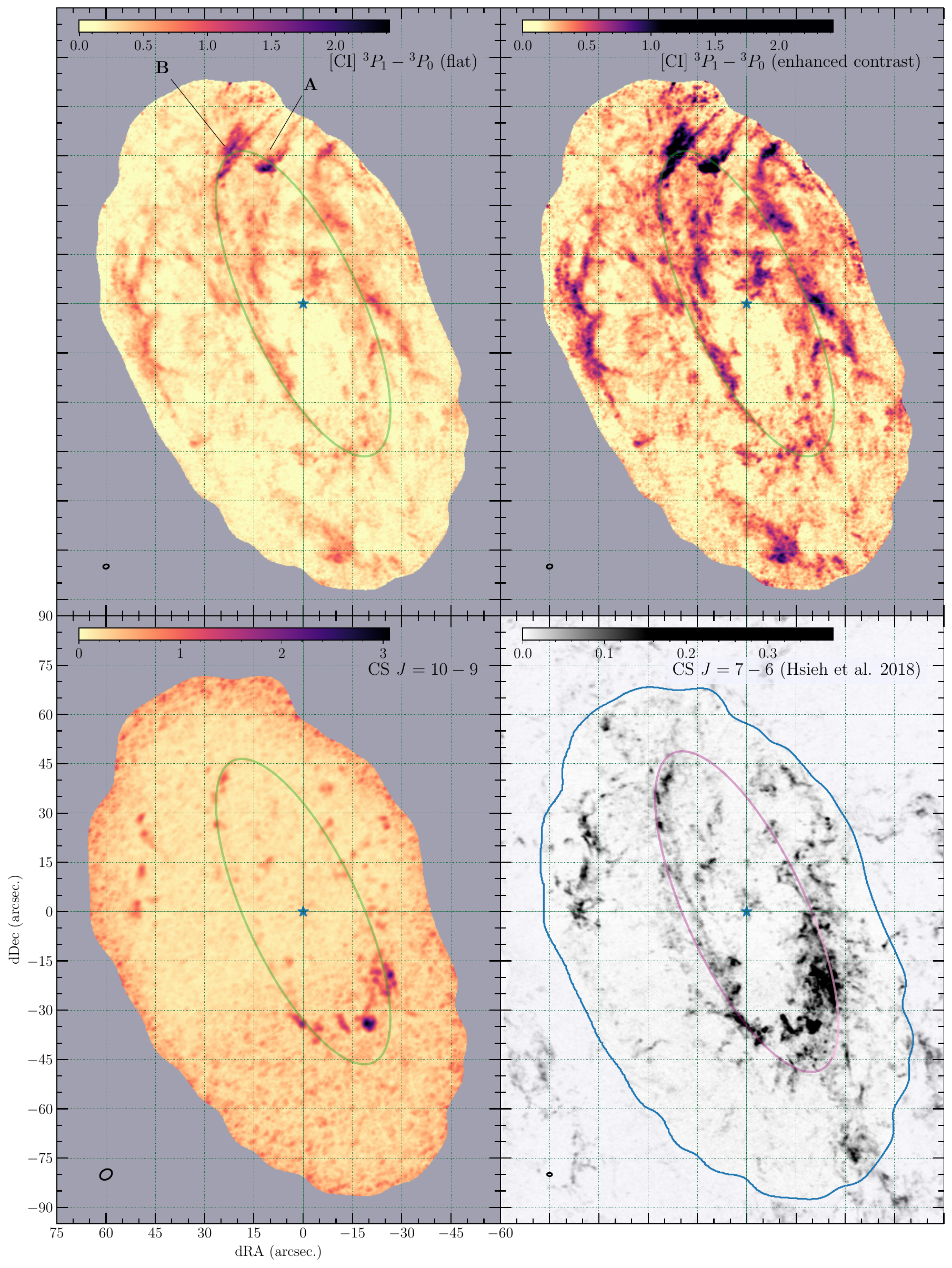}
\end{center}
  \caption{ Peak flux density maps of \CI~\CIa\ (upper left and right) and CS~\JJ{10}{9}\ (lower left).   The \CI\ image is shown in two different color scales.   The CS~\JJ{7}{6}\ map \citep[lower right; ][]{Hsieh2019} is shown for comparison.  Peak flux densities are calculated in 10-\kmps\ velocity bins.  The overlaid ellipse represents the approximate shape of the CND ring, based on parameters $\theta_0=\ang{25;;}, i=\ang{70;;}$ \citep{Jackson1993}, and a radius of 2.0~pc. \label{fig:maps}}.
  \end{figure*}

Figure~\ref{fig:maps} presents images of the peak flux densities of \CI~\CIa\ and CS~10--9.
The \CI\ map is displayed using two different color scales: a standard (flat) scale and an enhanced-contrast scale that emphasizes low-level emission.
For comparison, the figure also includes the CS7--6 image from \citet{Hsieh2019}.

Figure \ref{fig:maps} reveals a complex and widespread spatial distribution of \CI\ compared to the dense-gas distribution represented by CS~7--6.
The most prominent \CI\ features are two exceptionally bright clumps near the northern end of the CND ring, which are labeled A and B in the flat-scale \CI\ image. 
They exhibit \CI\ intensities approximately twice the maximum of the remaining \CI\ emission, but lack evident CS~7--6 counterparts. 
We refer to clumps A and B as the \CI-bright clumps (CBCs) hereafter.
In addition to them, numerous small \CI-bright clumps and filaments are widespread both inside and outside the CND ring, where CS~7--6 emission is only sparsely detected.

\begin{figure*}
\epsscale{1.2}
\plotone{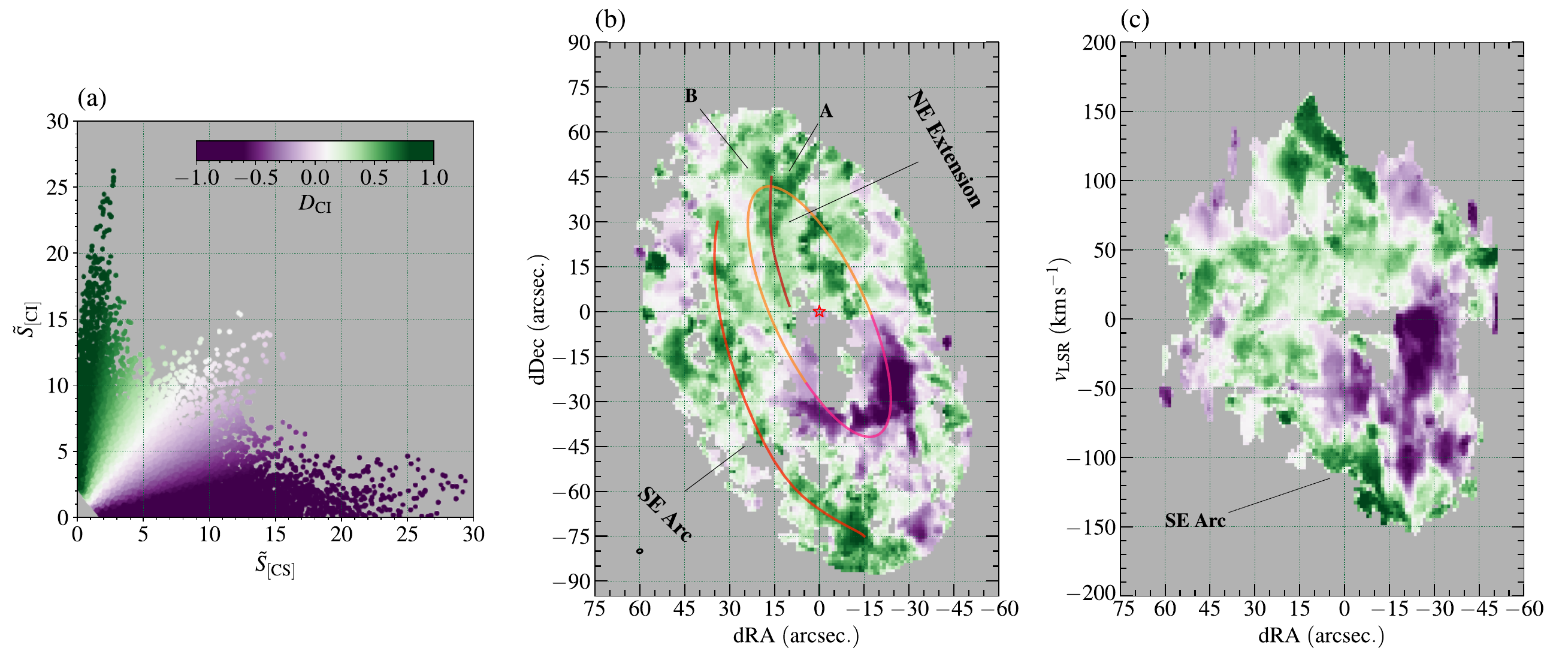}
\caption{(a) Scatter plot of \SCI\ vs. \SCS\ with spaxels colored by \Dci.  (b) \Dci\ distribution projected onto the \dRA--\dDec\ plane.  (c) \Dci\ distribution projected onto the \dRA--\vlsr\ plane.  The same color scale is used for all panels.\label{fig:DciMaps}}
\end{figure*}

To analyze the difference between the \CI\ and CS~7--6 distributions systematically, we introduce the \Dci\ parameter representing normalized intensity difference:
\begin{eqnarray}
\Dci &\equiv & \frac{\tilde{S}(\mathrm{CI}) - \tilde{S}(\mathrm{CS})}{\tilde{S}(\mathrm{CI}) + \tilde{S}(\mathrm{CS})},
\end{eqnarray}
where \SCI\ and \SCS\ are the \CI\ and CS~7--6\ intensities normalized by their median absolute deviations (MADs) from zero, respectively.
The \Dci\ values are calculated spaxel-wise in $\ang{;;1}\times\ang{;;1}\times{2.5~\kmps}$ \PPV\ bins.
\myrevision{The MAD values are measured for the spaxels with either of \CI\ or CS~7--6 qualifies 3-$\sigma$\ cutoff, so that they approximately represent the intensity scales for significant emission, not the noise levels.}
The MAD value is 0.11~\jkpb\ for \CI\ and 0.012~\jkpb\ for CS~7--6 in their original beam sizes.
We use \Dci\ instead of the raw intensity ratio because the poor spatial correlation between the two transitions causes the ratio to diverge to $\infty$ for a significant spaxels.
Figure \ref{fig:DciMaps} shows the \Dci\ distribution in the \SCI~vs~\SCS\ scatter plot, position--position (\PP) projection, and position--velocity (\PV) projection along the RA axis.
The scatter plot shows only a weak correlation between \SCI\ and \SCS; a minor portion of spaxels have neutral \Dci\ values of $\sim 0$, while the rest majority exhibit a polarized \Dci\ distribution into \CI-dominant (i.e., $\Dci\sim1$) and CS-dominant (i.e., $\Dci\sim-1$) regimes.   

The CS-dominant spaxels are almost exclusively found in the southern half of the CND ring.
In contrast, \CI-dominant spaxels are primarily distributed on and within the northern half of the ring, and outside the ring.
The most prominent \CI-dominant feature along with the two CBCs is a curved filament structure of \CI-dominant spaxels, which extends from $(\dRA, \dDec) \sim (\ang{;;30},\ang{;;30})$ to $\sim (\ang{;;0},\ang{;;-70})$.
We label this structure as the southeastern (SE) arc in Figure \ref{fig:DciMaps}.
The SE filament is not clearly discernible in the original \CI\ or CS images but appears as a partial ring of high-\Dci\ spaxels surrounding the CND ring at approximately twice the radius.
The \PV\ image indicates that the SE arc has systematically higher \vlsr\ than the CND ring, suggesting faster circular velocities or the presence of inward radial motion.
Another remarkable \CI-dominant feature is a streamer-like structure inside the CND, extending from the northern end of the ring toward \sgras.
We label this feature the northeast (NE) extension, following the nomenclature in \cite{Christopher2005}.
The NE extension in the \CI\ image appears to extend further into the inner part of the cavity than in dense-gas images, reaching a distance of $\sim 0.5$~\pc\ from \sgras\ in projection.

\begin{figure*}
\epsscale{0.95}
\plotone{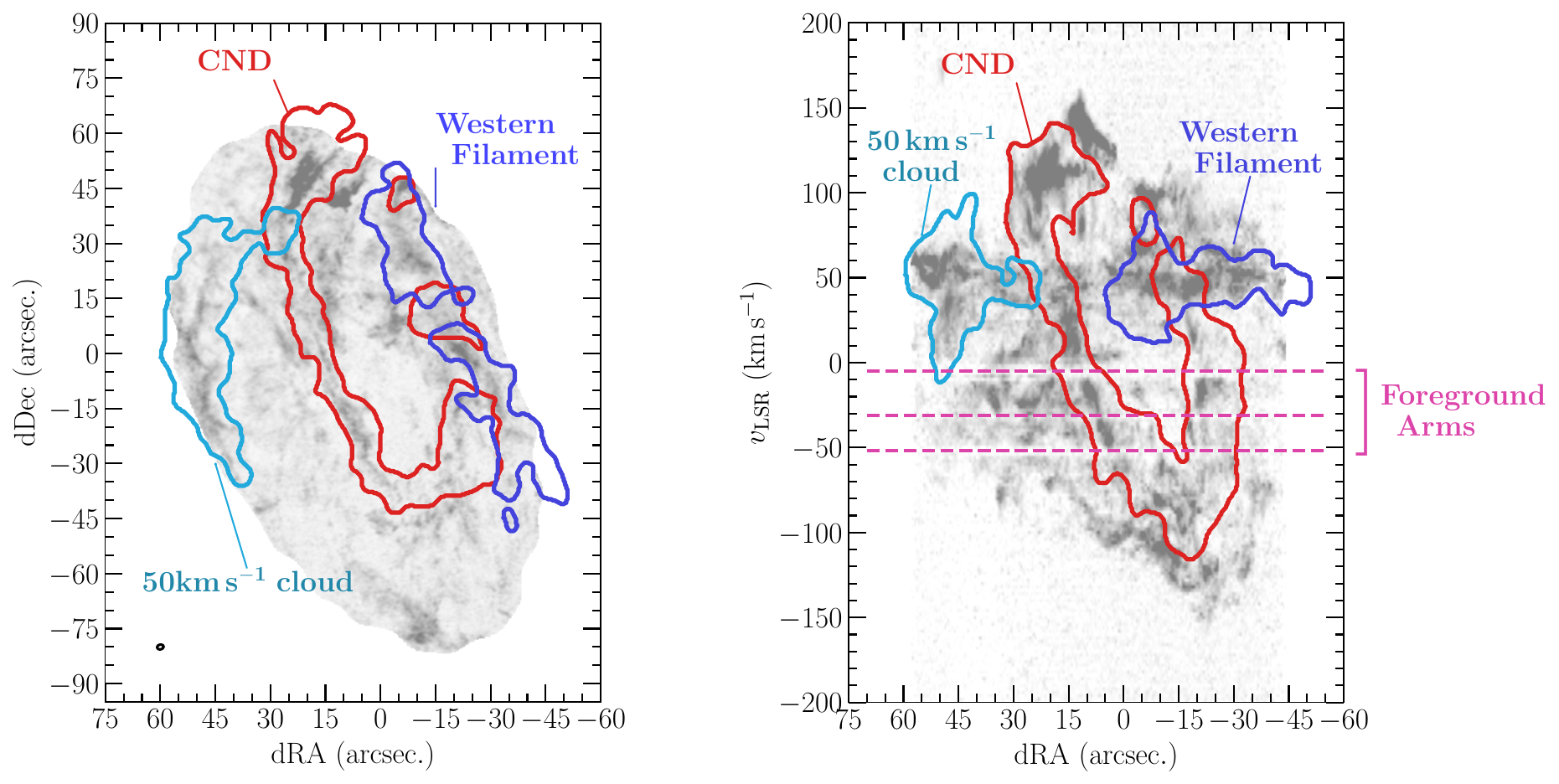}
\caption{ Non-CND and candidate non-CND features on \CI\ peak intensity image (left) and the \PV\ diagram projected along the RA axis (right). The solid lines indicate approximate shapes of the CND ring, 50-\kmps\ cloud, and Western Filament.  The dashed lines in the \PV\ diagram mark absorption bands of the foreground spiral arms.  \label{fig:SnormMaps} }
\end{figure*}

\subsection{CS 10--9}
The overall CS~10--9 distribution is consistent with the CS~7--6 image, with bright emission predominantly associated with the southern rim of the CND ring.
The intensity contrast between the northern and southern portions of the ring is more pronounced in the CS~10--9 image compared to CS~7--6.
This suggests a concentration of the highest-density gas toward the southern portion of the ring, which roughly corresponds to the southern extension and southwestern (SW) lobe \citep{Christopher2005}.
However, since this paper focuses on low-density molecular gas, we do not investigate the details of the density structure of the dense-gas component further.

\subsection{Non-CND features \label{subsection:results:nonCNDfeatures}}

Figure~\ref{fig:SnormMaps} shows the \dRA--\dDec\ and \dRA--\vlsr\ projections of the \CI\ data.
The \PV\ diagram reveals a horizontal feature at $50~\kmps$ that spans the entire RA range.
This feature is separated into eastern and western components in the \PP\ image.
The eastern component is spatially detached from the CND and offset by \ang{;;30} toward the 50-\kmps\ cloud, indicating that it is the western edge of the 50-\kmps\ cloud, which lies just outside the FoV.
The western component, labeled the ``Western Filament'' in the figure, lies on the opposite side of the 50-\kmps\ cloud, spatially coinciding with the western rim of the CND.
This spatial relationship suggests that the Western Filament might be a CND-associated feature.
However, we conservatively interpret it as part of the 50-\kmps\ cloud based on its centroid velocity and lack of a velocity gradient indicative of rotating motion.

Three narrow absorption features at $\vlsr = 0$, $-30$, and $-50$~\kmps\ in the \PV\ diagram are attributed to foreground spiral arms.
The 50-\kmps\ cloud, Western Filament, and foreground spiral arms are excluded from the remainder of the analysis in this paper.

\subsection{Rings and Streamers\label{subsection:results:3Dkinematics}}

\subsubsection{0- and 80-KM anomalies \label{subsubsection:results:3Dkinematics:0and80kmComp}}
\begin{figure*}
\epsscale{.925}
\plotone{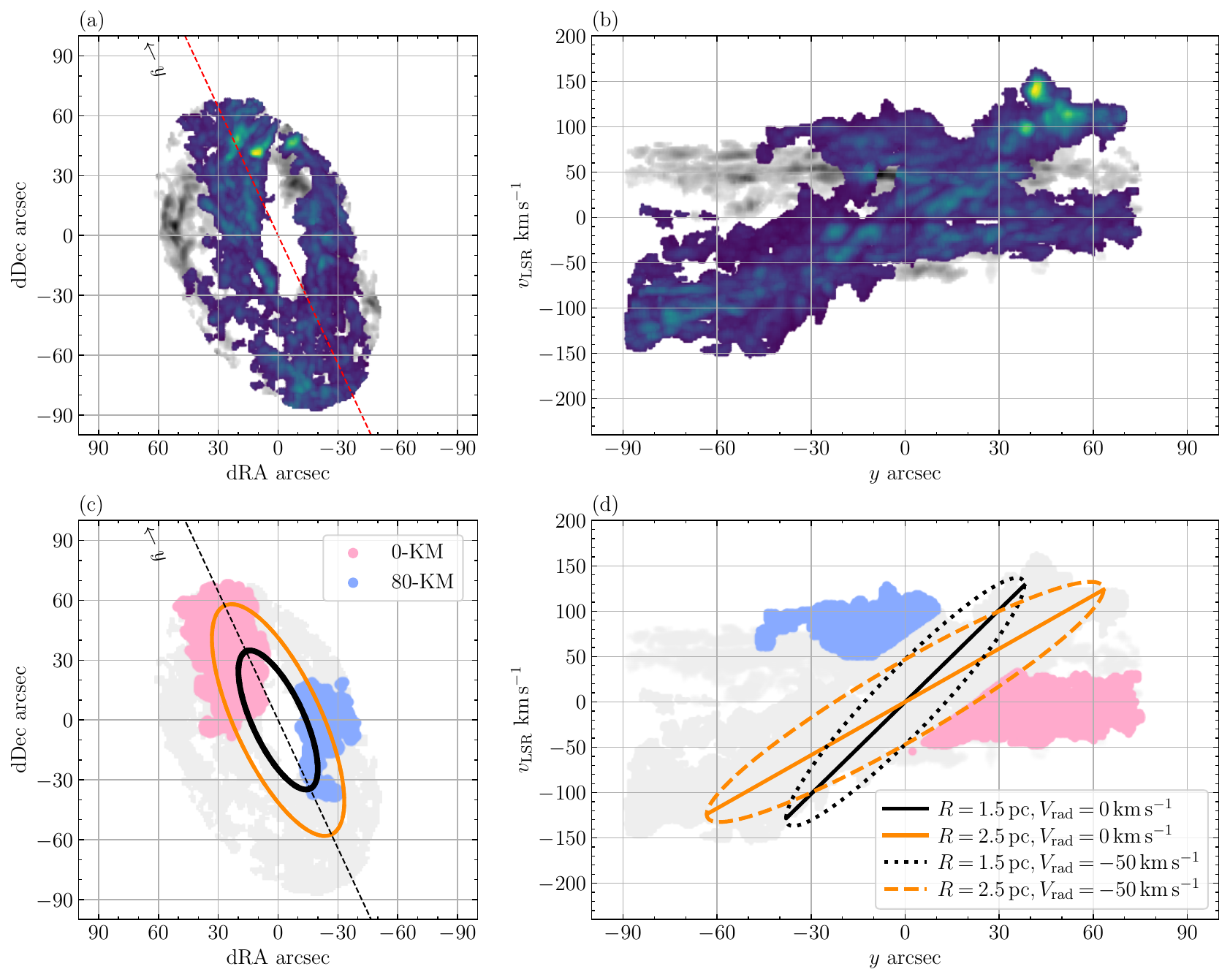}
\caption{(a) \CI\ peak intensity on the \alpIRCS--\deltIRCS\ plane. (b) Position--\vlsr\ diagram along the inclination axis ($y$) indicated in panel (a). Non-CND emission \myrevision{identified in \S\ref{subsection:results:nonCNDfeatures}} is shown in grayscale in both panels. (c, d) \myrevision{Same as (a) and (b), but with the 0-KM and 80-KM anomalies highlighted in red and blue, superposed on other CND and non-CND emissions in gray}.  Overlaid model curves are for $R = 1.5$~pc and 2.5~pc and radial velocities \Vrad\ = 0~\kmps\ and $-50$~\kmps. \label{fig:0and80kmComp}}
\end{figure*}

\myrevision{
Previous studies (e.g., \citealt{Jackson1993}; \citealt{Hsieh2015}; \citealt{Tsuboi2018}) have consistently shown that the three-dimensional kinematics of the molecular gas associated with the CND can be represented by a purely rotating ring, streamers rotating while inflowing with a finite radial velocity (\Vrad), or a combination of both.
In this subsection, we describe the CND system as consisting of multiple rotating and streaming components on one consistent orbital plane and decompose them by calculating \Vrad\ on a spaxel-by-spaxel basis.
}

However, the observed \CI\ emission contains velocity anomalies that deviate significantly from the CND's kinematics, which must be treated separately from the primary components.
Figure \ref{fig:0and80kmComp} shows the \CI\ distribution in the \dRA--\dDec\ and $y$--\vlsr\ projections, where $y$ denotes the position along the inclination axis.
Model curves for a rotating ring at radii of 1.5~pc and 2.5~pc are overlaid, alongside curves with a uniform radial \Vrad\ of $-50$~\kmps\ at the same radii.
The \PV~diagrams show two velocity components without significant velocity gradients that cannot be fitted with rotating or streaming models regardless of the assumption for \Vrad.  
We refer to them as the 0-KM\ and 80-KM anomalies according to their representative \vlsr.

The 80-KM anomaly approximately corresponds to the dense-gas feature Anomaly-A \citep{Tsuboi2018}.
The 0-KM anomaly resembles the local arm in the centroid velocity of $\sim 0$~\kmps\ and the absence of a velocity gradient.
However, its broad velocity extension of $\sim50$~\kmps\ clearly distinguishes it from the spiral arms, which typically exhibit velocity widths of a few \kmps.
We exclude these two anomalies, along with the non-CND features identified in \S\ref{subsection:results:nonCNDfeatures}, from the $R$--\Vrad\ diagram analysis described later in this subsection.
These anomalies are likely to be either foreground/background CMZ clouds similar to the 50-\kmps~cloud, or CND-associated clouds moving on a different plane from the CND, as we discuss in detail later (\S\ref{subsection:results:anomalies}).


\subsubsection{R--\Vrad\ Diagram\label{subsubsection:results:3Dkinematics:RVdiagram}}

\begin{figure*}
    \epsscale{1.}
    \plotone{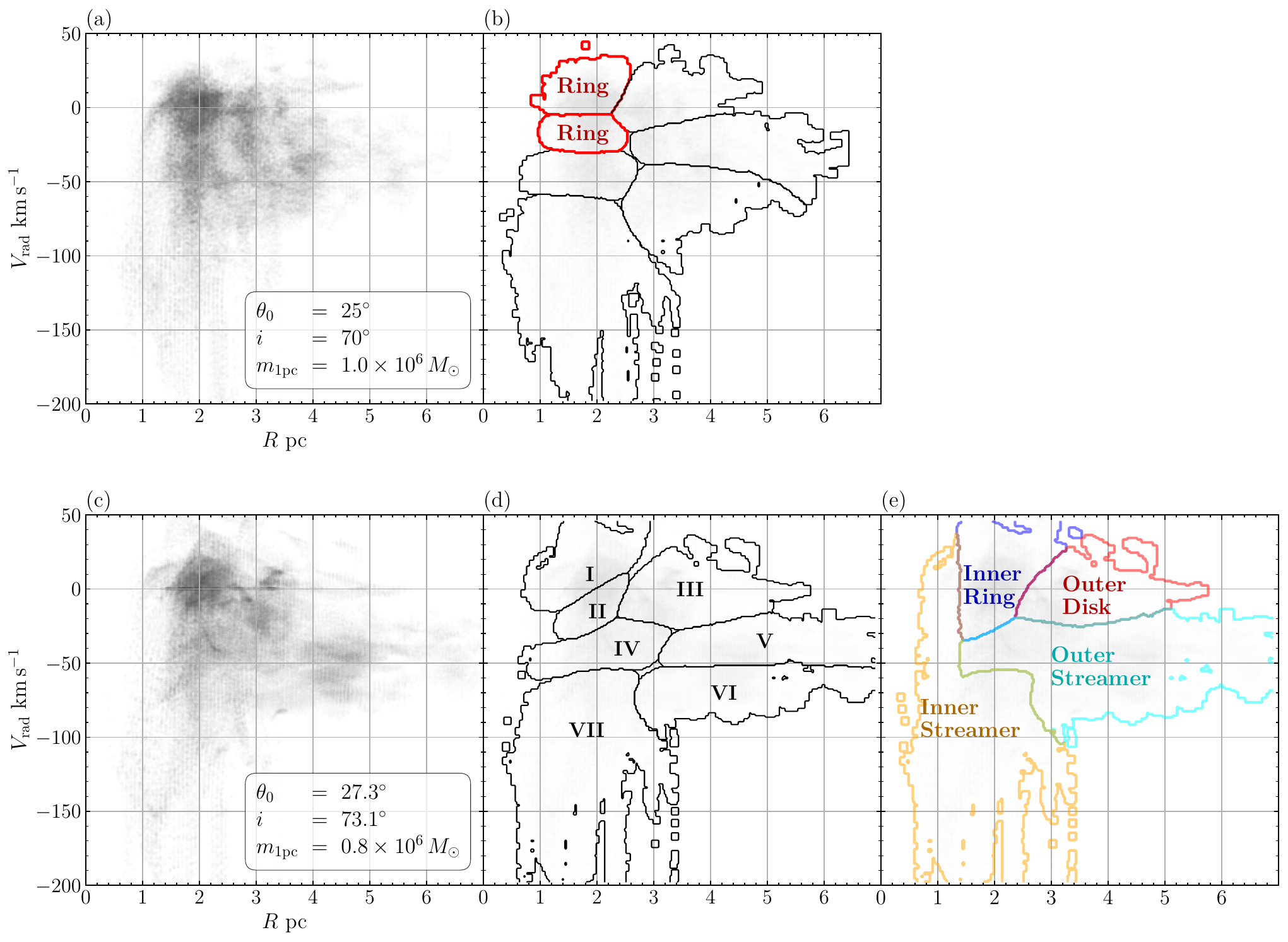}
    \caption{\myrevision{Results of the $R$--\Vrad\ diagram analysis. (a) Initial $R$--\Vrad\ diagram. (b) Raw SMM decomposition of panel (a). (c, d) Same as (a, b), respectively, but showing the final iteration results. (e) Definition of the four subregions in the $R$--\Vrad\ diagram.\label{fig:mgmgres}}}    
\end{figure*}

\begin{figure*}
    \epsscale{1.0}
    \plotone{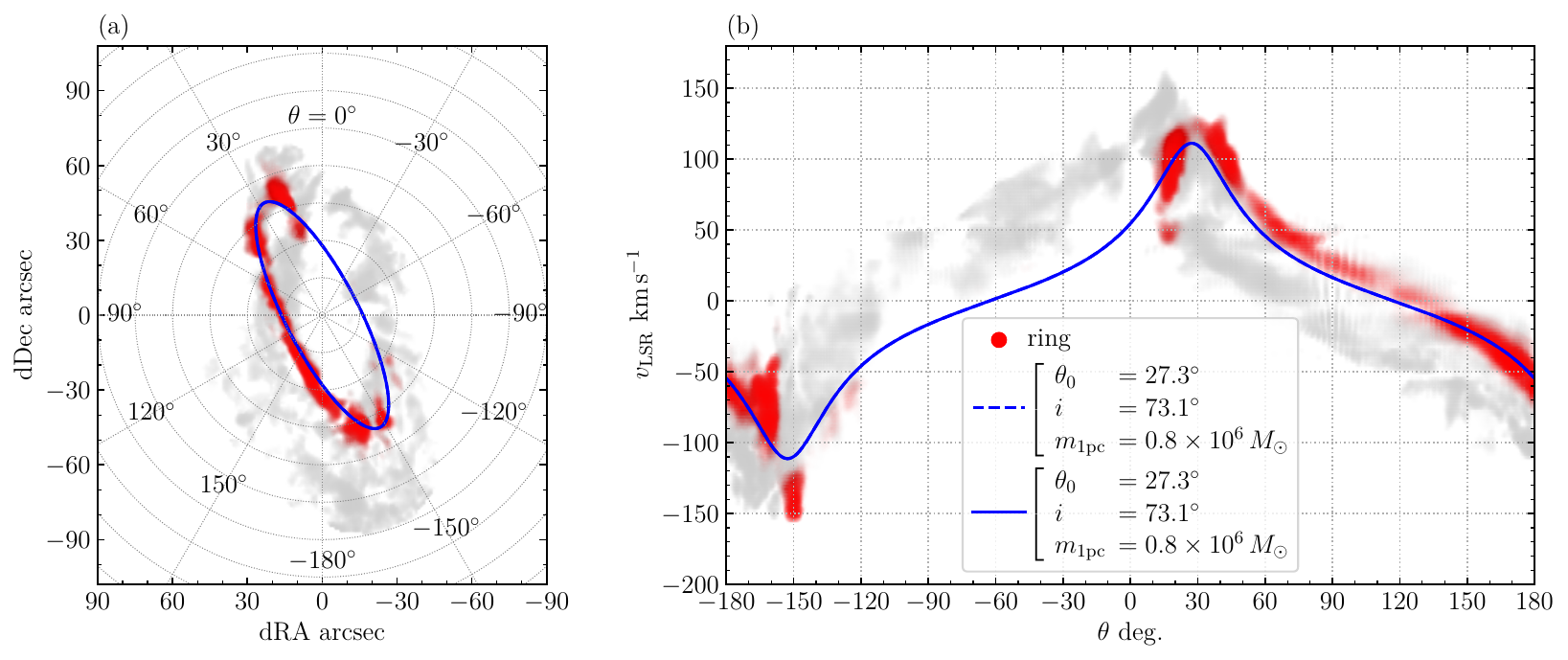}
    \caption{\myrevision{Purely rotating ring models overlaid on (a) the \CI\ image and (b) the $\theta$--\vlsr\ phase diagram. Dashed curves represent the initial model parameters ($\theta_0 = \ang{25;;}$, $i = \ang{70;;}$, $\mstar = 1.0 \times 10^6~\Msun$), and solid curves represent the best-fit parameters ($\theta_0 = \bestAlp$, $i = \bestI$, $\mstar = 0.8 \times 10^6~\Msun$). Spaxels used for the orbit fitting are shown in red.\label{fig:ofit}}}    
\end{figure*}

\begin{figure}
    \epsscale{1.0}
    \plotone{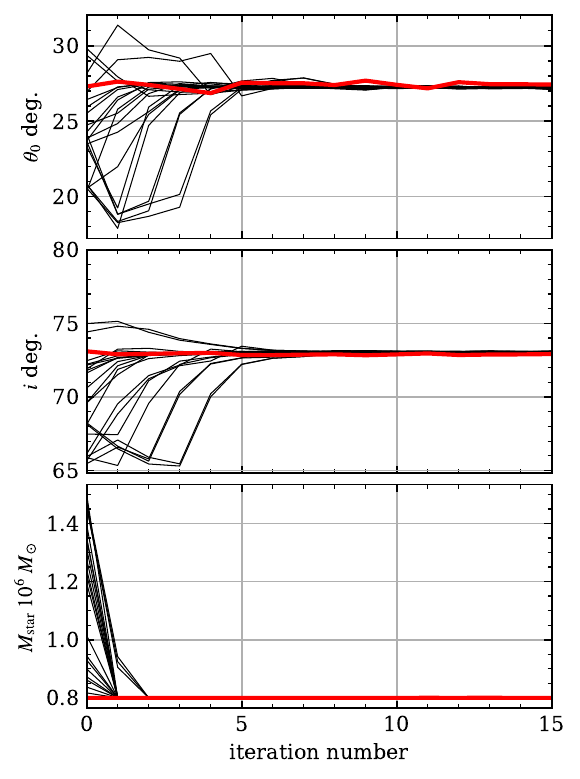}
    \caption{\myrevision{Trace plots of $\theta_0$, $i$, and \mstar\ over 15 iterations.  The red lines are for the main analysis starting from $\theta_0 = \ang{25;;}$, $i = \ang{70;;}$, and $\mstar = 0.8\times10^6~\Msun$, whereas others  show the parameters starting from randomly chosen initial values.  \label{fig:trasceplot}}}    
\end{figure}

In the following, we analyze the three-dimensional kinematics of the CND and CND-associated features by projecting the \CI\ spaxels onto the orbital plane coordinate defined by radius ($R$), PA ($\theta$), and \Vrad.
This conversion requires a radial mass distribution model $M\left(R\right)$ upon which the circular velocity $\Vrot= \sqrt{\frac{GM\left(R\right)}{R}}$ is calculated.  
We adopt the model based on the density distribution by \cite{Schodel2018} :
\begin{eqnarray}
M\left(R\right) &=& m_\mathrm{BH} + \mstar\cdot\left(\frac{R}{1~\mathrm{pc}}\right) ^ {\left(-\gamma+3\right)}, \label{eqn:enclosedMass}
\end{eqnarray}
where \mbh\ and \mstar\ are the \sgras\ mass ($4.1\times10^6~\Msun$; \citealt{GravityCollaboration2019}) and the stellar component mass within $R\leq1~\pc$, respectively, and $\gamma = 1.13$ \citep{Schodel2018}.
The measured values of \mstar\ range from $1\times10^6~\Msun$ to $2\times10^6~\Msun$ (\citealt{Schodel2018}).  We treat \mstar\ as a variable parameter constrained in the range $\left(0.8\mbox{--}2\right)\times10^6~\Msun$, which is slightly wider than the observed range.

\myrevision{
The orientation of the CND’s orbital plane, defined by the position angle ($\theta_0$) and inclination ($i$), must be determined to deproject the plane-of-the-sky coordinates into three-dimensional positions.
As noted at the beginning of this section, we assume that the \CI\ emission consists of a mixture of purely rotating and streaming components.
We identify spaxels near $(R, \Vrad) \sim (2~\pc, 0~\kmps)$ on the $R$--\Vrad\ diagram as the purely rotating ring family and derive the orbital plane parameters by fitting them with a rotating-ring model.
We adopt a self-consistent approach in which the fitting of the orbital plane and the coordinate deprojection are iteratively refined.
Each iteration cycle comprises three steps:
}
\begin{enumerate}
\item Coordinate conversion from (\alpIRCS, \deltIRCS, \vlsr) to ($R$, $\phi$, \Vrad) for \myrevision{all purely rotating and streaming spaxels}, where ($\alpIRCS$, $\deltIRCS$) is the relative R.A.--Dec. position from \sgras.  This step requires assumptions for $\theta_0$, $i$, and \mstar.
\item Identification of the purely rotating spaxels on the $R$--\Vrad\ diagram with the Student-$t$ mixture model (SMM).
\item Refinement of $\theta_0$, $i$, and \mstar\ by fitting the purely rotating spaxels identified at Step 2 with the rotating ring model.
\end{enumerate}
Steps 1--3 are repeated until $\theta_0$, $i$, and \mstar\ converge, starting with the initial parameters  $\theta_0 = \ang{25;;}$, $i = \ang{70;;}$ \citep{Jackson1993} and \mstar = \myrevision{$1\times10^6$}~\Msun.

The two coordinate systems (\alpIRCS, \deltIRCS, \vlsr) and ($R$, $\phi$, \Vrad) are related through the following 
 equations: 
\begin{eqnarray}
\sqrt{\alpIRCS^2 + \deltIRCS^2} &=& \frac{R}{D} \sqrt{\left(\sin\phi\cdot\cos i\right)^2 + \cos^2\phi}, \label{eqn:proj1} \\
\arctan\left(\frac{\deltIRCS}{\alpIRCS}\right) &=& \arctan\left(\frac{\sin\phi\cdot\cos i}{\cos\phi}\right) + \theta_0   \label{eqn:proj3} ,\\
v_\mathrm{lsr} &=& \left\{V_\mathrm{rad}\sin\phi + \Vrot\left(R\right) \cos\phi\right\}\sin i, \label{eqn:vrad}
\end{eqnarray}
where \Vrad\ is defined as positive for outward motion.
The solution of Equations \ref{eqn:proj1}--\ref{eqn:vrad} becomes highly unstable near the inclination axis, where the contribution of \Vrad\ to \vlsr\ diminishes as $\sin\phi\rightarrow 0$.
To address this issue, we adopted a Bayesian approach with a smoothness prior instead of straightforwardly solving the equations.
Details of this analysis are provided in Appendix \ref{appendix:fitting}.

\begin{revision}
Equations~\ref{eqn:proj1} and \ref{eqn:proj3} omit the vertical extent ($\Delta z$) from the orbital plane, which can lead to potential misinterpretation from $\left(\alpha, \delta\right)$ to $\left(R, \phi\right)$, as $\Delta z$ may have a finite value comparable to the width of the CND ring (i.e., $\sim0.3~\pc$).
The resulting uncertainty in \Vrad\ due to $\Delta z$, propagated through Equations~\ref{eqn:proj1}--\ref{eqn:vrad}, can be approximated as $\sim\frac{1}{2}\tan i \cdot \sin 2\phi \cdot \frac{\Vrot(R)\cdot\Delta z}{R}$, where the secondary term arising from the weak $R$-dependence of $\Vrad(R)$ within the analyzed $R$ range is ignored.
This approximation predicts enhanced dispersion in the $\phi$--\Vrad\ curve around $\pm\ang{45;;}$ and $\pm\ang{135;;}$ if $\Delta z$ significantly affects the \Vrot\ estimate.
However, we confirm that no such signature is observed in the best-fit $\phi$--\Vrad\ curve presented in Appendix~\ref{appendix:fitting}, indicating the effect of $\Delta z$ can be safely ignored.
\end{revision}

Figure \ref{fig:mgmgres}a shows the $R$--\Vrad\ diagram with the initial parameters. 
The diagram is decomposed into multiple clusters by employing SMM, which fits the density distribution of data points with a superposition of multiple Student-$t$ kernels.   \myrevision{An outline of the SMM is provided in Appendix \ref{appendix:SMM}.}
The shape parameter $\nu$ of the Student-$t$ kernel is chosen to be 4.
The number of clusters is fixed at 7, which is the largest number that provided consistent decomposition with varying $\theta_0$, $i$, and \mstar\ throughout the entire iterations.
\myrevision{We excluded outliers on the $R$--\Vrad\ plane with $\Vrad < -150~\kmps$ from the SMM decomposition to stabilize the iteration; without this additional condition, the iteration process failed to converge for several initial parameter sets.}
The SMM results are overlaid on the initial diagram in Figure \ref{fig:mgmgres}b.
Note that SMM does not yield strict boundaries between clusters; one data point can belong to multiple clusters, according to the responsibility $\gamma_{mn}$ ranging [0 1], which represents the probability that the $m$th spaxel belongs to the $n$th cluster.
For simplicity, boundary lines are drawn where $\gamma_{mn} = 0.5$.
The two clusters around $(R, \Vrad) = (0~\pc, 0~\kmps)$, which are marked red in the figure, represent the purely rotating ring spaxels.  Their \PPV\ positions are shown in Figure \ref{fig:ofit}.

After defining the purely rotating ring spaxels, we refine the parameters $\theta_0$, $i$, and \mstar\ by fitting these spaxels with the steadily rotating ring model, represented by Equations \ref{eqn:proj1}--\ref{eqn:vrad}, with \Vrad\ fixed at 0~\kmps\ and the ring radius treated as a free parameter.
The fitting results from the first iteration are shown in Figure~\ref{fig:ofit}.
The self-consistent parameters obtained after iteration are \myrevision{$\theta_0 = \bestAlp$, $i = \bestI$}, and $\mstar = 0.80\times10^6~\Msun$.
We adopt these values of $\theta_0$, $i$, and \mstar\ throughout the remainder of the paper unless stated otherwise.
\begin{revision} 
We also performed 20 analysis runs starting from parameters randomly selected within the ranges
$\theta_0 \in \left[\ang{20;;}, \ang{30;;}\right]$,
$i \in \left[\ang{65;;}, \ang{75;;}\right]$, and
$\mstar \in \left[0.8\times10^6~\Msun, 1.5\times10^6~\Msun\right]$.
Figure~\ref{fig:trasceplot} shows the trace plots of the parameters against iteration number, confirming that they converge to unique values regardless of the initial conditions.
The standard deviations of the convergence values among the different runs are $\sigma_{\theta_0} = \ang{0.08;;}$, $\sigma_i = \ang{0.05;;}$, and $\sigma_{\mstar} = 4~\Msun$. Note that \mstar\ is determined by a lower-limiting condition, resulting in an almost zero $\sigma_{\mstar}$.
The convergence values also weakly depend on the analysis setup, such as the choice of parameter ranges, hyperparameters (see Appendix~\ref{appendix:fitting}), and the cutoff \Vrad\ level in the SMM decomposition. This arbitrariness affects $\theta_0$ and $i$ by approximately $\ang{1;;}$.
We confirmed that this level of uncertainty is small enough that the subsequent analysis remains essentially unaffected.
\end{revision}

\begin{figure*}
\epsscale{1.}
\plotone{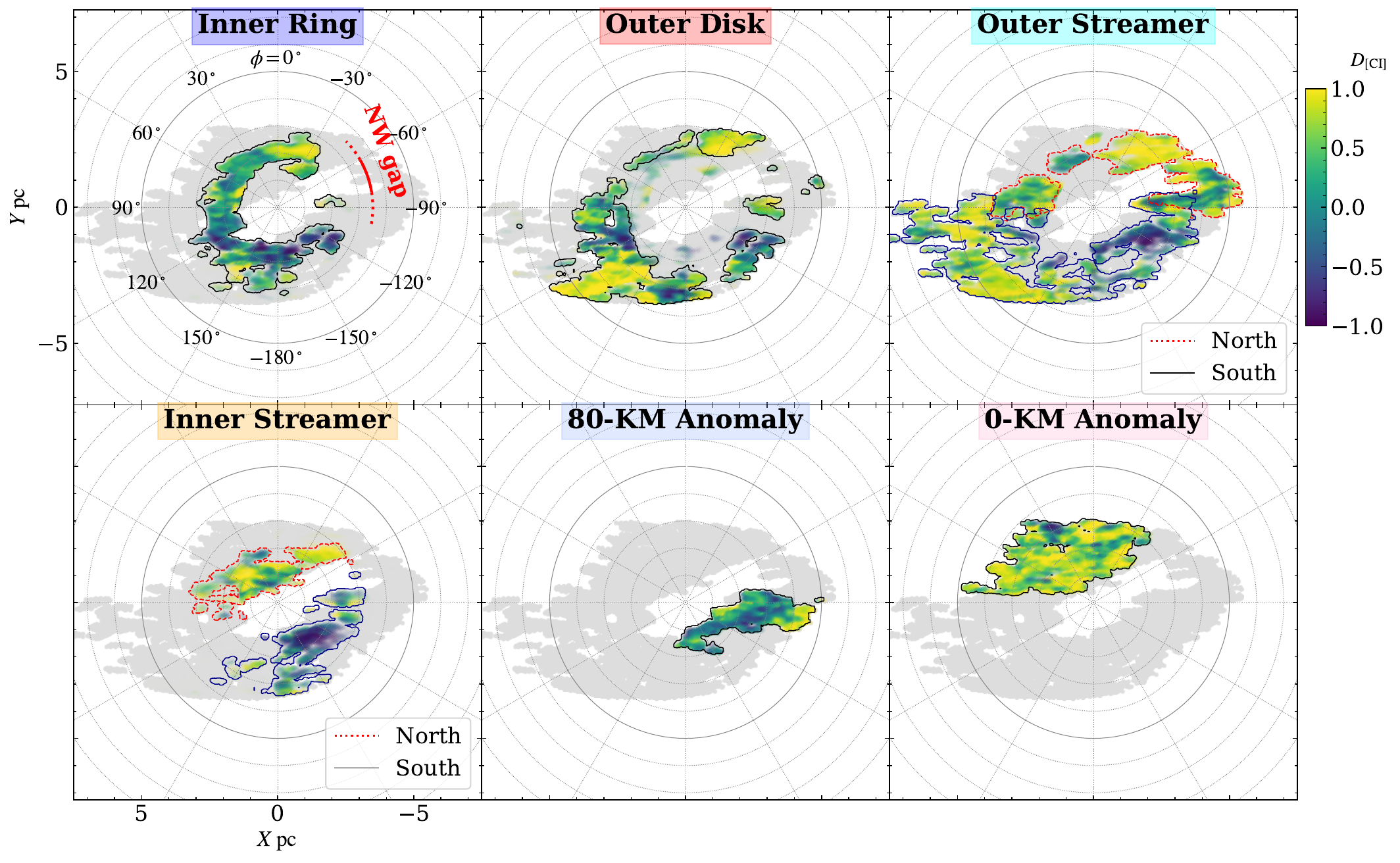}
\caption{\myrevision{Subregions on the $X$--$Y$ plane, with the color representing \Sdense, overlaid on the total \CI\ emission shown in gray.} \label{fig:regPP}}
\end{figure*}

The $R$--\Vrad\ diagram and the model curves with the final parameters are shown in Figures~\ref{fig:mgmgres}c,d and \ref{fig:ofit}, respectively.
The entire diagram can be roughly divided into four sections.
The first consists of small-$R$ and low-\Vrad\ spaxels, corresponding to the purely rotating ring around $(R, \Vrad) \sim (0~\pc, 0~\kmps)$, which is more pronounced than in the initial diagram.
The second includes a large-$R$, high-\Vrad\ region with a distinct aggregation centered at $(R, \Vrad) \sim (4.5~\pc, -50~\kmps)$, interpreted as inflowing material outside the CND.
The third is a population of large-$R$, low-\Vrad\ spaxels forming a horizontal structure spanning $R \sim 3$--6~\pc\ at $\Vrad \sim 0$~\kmps, appearing as an outer extension of the purely rotating ring.
The fourth consists of small-$R$, high-\Vrad\ spaxels at $R \lesssim 2$~\pc, including a nearly vertical structure at $R \sim 1$~\pc\ and $\Vrad \sim -150$ to $-50$~\kmps. 
This feature likely traces inflowing material from the CND toward the central cavity.
We reorganize the seven SMM-identified clusters I--VII shown in Figure~\ref{fig:mgmgres}d into four kinematic components to match these four sections, which we label as the Inner Ring, Outer Streamer, Outer Disk, and Inner Streamer in the figure.
Their boundaries are shown in \ref{fig:mgmgres}e.
The spatial ($X$--$Y$) distributions of the four components are presented in Figure \ref{fig:regPP}, where $\left(X, Y\right) = \left( R\cdot\sin\phi, R\cdot\cos\phi\right)$ is the coordinate on the orbital plane. 
The \Dci\ values are represented by colors in each panel.
The following describes each component individually.

\subsubsection{Inner Ring (CND-IR) \label{subsubsection:results:3Dkinematics:IR}}

The Inner Ring (CND-IR) consists of the spaxels associated with the primary aggregation of the $R$--\Vrad\ diagram at $(R, \Vrad) \sim (2~\pc, 0~\kmps)$.
We define the $R \geq 1.4~\pc$ portion of regions I and II as CND-IR.
The emission in the $R < 1.4~\pc$ region is excluded, as it lies well inside the CND ring in the raw \CI\ image and has a \PPV\ distribution that is more continuous with the Inner Streamer described in \S\ref{subsubsection:results:3Dkinematics:IS}. 

CND-IR approximately corresponds to the CND ring, except for the large \Vrad\ portion in the SW lobe being assigned to CND-IS.
Its \PPV\ distribution is consistent with the circular ring of a $2.0$~\pc\ radius with $i= \bestI$ and $\theta_0=\bestAlp$ rotating counterclockwise at $\Vrot$ of $125$~\kmps.
A slight offset from the purely rotating model in the $\theta$--\vlsr\ phase plot (Figure \ref{fig:ofit}b) suggests that CND-IR may have a small inward velocity.
The median \Vrad\ weighted by the \CI\ intensity in the $R$--\Vrad\ diagram is $-3.7~\kmps$.

The ring has an emission gap in the northwestern portion ($\phi\sim$\ang{-120;;} to \ang{-30;;}), which is consistently present in known CND images.
We refer to it as the northwest (NW) gap hereafter.
The southern half of the ring is CS-dominant, while \CI\ becomes increasingly dominant as $\phi$ decreases clockwise until $\Dci\sim1$ at the northern edge of the NW gap ($\phi\sim\ang{-30;;}$).
Thus, the NW gap represents a discontinuity in both the \Dci\ variation and the intensity distribution.

\subsubsection{Outer Disk (CND-OD)  \label{subsubsection:results:3Dkinematics:OD}}
The Outer Disk (CND-OD) corresponds to cluster~II, which appears as a horizontal structure at $R = 3$--6~\pc\ and $\Vrad\sim0$~\kmps\ in the $R$--$\Vrad$ diagram.
The \CI-weighted median $\Vrad$ is $-0.5$~\kmps.
Although we refer to this component as a ``disk,'' the \CI\ emission does not uniformly fill the 3--6~\pc\ radial range; rather, it is unevenly distributed with respect to $\phi$, exhibiting its largest radial extent around $\phi = \ang{120;;}$--$\ang{150;;}$.
CND-OD also contains more \CI-dominant spaxels than CND-IR, particularly in the $\phi$ ranges of \ang{120;;}--\ang{150;;} and \ang{-30;;}--\ang{30;;}, where CND-OD overlaps with CND-OS in projection.
This weakness in dense-gas tracer emission, along with its irregular morphology, likely contributed to CND-OD being indistinct in previous molecular line images.

\subsubsection{Outer Streamer (CND-OS)  \label{subsubsection:results:3Dkinematics:OS}}
The Outer Steamer (CND-OS) consists of clusters IV--\myrevision{VI}, which are associated with the secondary aggregation in the $R$--\Vrad\ diagram at  $(R, \Vrad) \sim (4.5~\pc, -50~\kmps)$.
CND-OS is the most radially extended component, whose $R$ ranges from 1.5~\pc\ to 7~\pc.  
CND-OS partially overlaps CND-IR and -OD in the radial distribution, but is clearly distinguished by the large negative \Vrad, whose \CI-weighted median is $-39.3$~\kmps.
The overall shape of the outer stream is an ellipse elongated in the $\phi\sim\ang{120;;}$ to $\sim\ang{-60;;}$ direction, with a pair of protrusions at $\phi\sim\pm\ang{90;;}$ where the radial extension exhibits local maxima. 
The entire structure could be described as a pair of streamers originating from these protrusions and spiraling inward in a counterclockwise direction.   
We define these two streamers as CND-OS south and north (denoted as CND-OS(N) and CND-OS(S), respectively), whose boundary is indicated in Figure \ref{fig:regPP}.
CND-OS(N) approximately corresponds to the dense-gas features called the Western streamer \citep{Hsieh2017} and Anomaly-B \citep{Tsuboi2018}.

CND-OS is more \CI-dominant than CND-OD, especially at its outer ($\gtrsim 4$~pc) radii.
The \CI-bright emission extended outside the CND ring is mostly contribution from CND-OS. 
The \CI-bright SE arc identified in the \Dci\ image (Figure \ref{fig:DciMaps}) is included in CND-OS(S).
The generally high \Dci\ indicates that CND-OS is more massive than previously considered from the dense-gas data.
The mass estimate based on the \CI\ and CS\ luminosities will be provided in \S\ref{subsection:results:massEstimate}.

\subsubsection{Inner Streamer (CND-IS)  \label{subsubsection:results:3Dkinematics:IS}}
The Inner Streamer (CND-IS) is the innermost and highest \Vrad\ component, consisting of cluster VII and the $R<1.4~\pc$ region of clusters I and II.  The \CI-weighted median \Vrad\ is $-68.5$~\kmps.  The majority of CND-IS lies inside the CND ring; in particular, the inner edge of the $\phi\sim$\ang{30;;}--\ang{90;;} portion extends to less than $R\sim 0.5$~pc from \sgras. 

CND-IS is irregularly shaped compared with the other three components.
Emission gaps at $\phi\sim\ang{125;;}$ and $\sim\ang{-70;;}$ separate CND-IS into the northern and southern parts, which we refer to CND-IS(N) and CND-IS(S), respectively.
CND-IS(N) corresponds to the NE extension, a \CI-bright feature identified in \S\ref{subsection:results:images}.
Contrastingly, CND-IS(S) contains the SW lobe where CS~7--6 and 10--9 are brightest but \CI\ is relatively weak.
The CND-IS is immediately adjacent to the northern arm (NA) and western arc (WA) of the mini-spiral.  Their spatial-velocity correlation is examined later in detail (\S\ref{subsection:results:RRLs}).
The 0-KM and 80-KM anomalies are also adjacent to CND-IS(N) and (S), respectively, suggesting possible physical continuity between them, though CND-IS and the anomalies differ in \vlsr\ by $\sim 50$~\kmps.
We will discuss this issue in \S\ref{subsection:results:anomalies}.

\subsection{Radial and Azimuthal \Dci\ Variations \label{subsection:results:CItoCSratio}}

\begin{figure*}
\epsscale{0.9}
\plotone{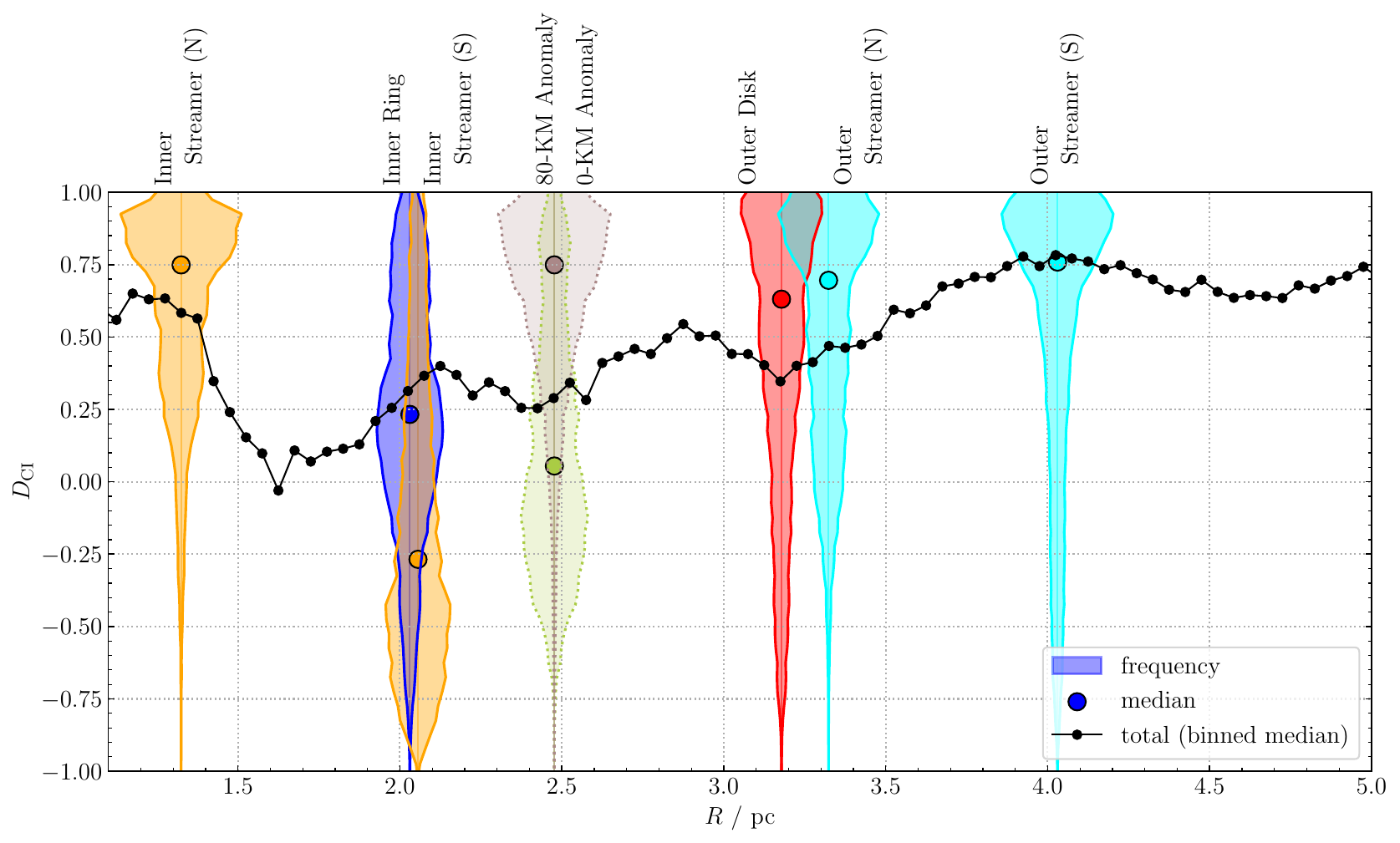}
\caption{
\myrevision{Radial \Dci\ variation.
Violin plots represent \Dci\ frequency histograms for individual components, with the horizontal position representing the median $R$. The radial variation of \Dci\ for the total CND-related emission is overlaid, with data points indicating the median \Dci\ within 0.05~\pc\ radial bins. \label{fig:Dci}}}
\end{figure*}

\begin{figure*}
\epsscale{0.95}
\plotone{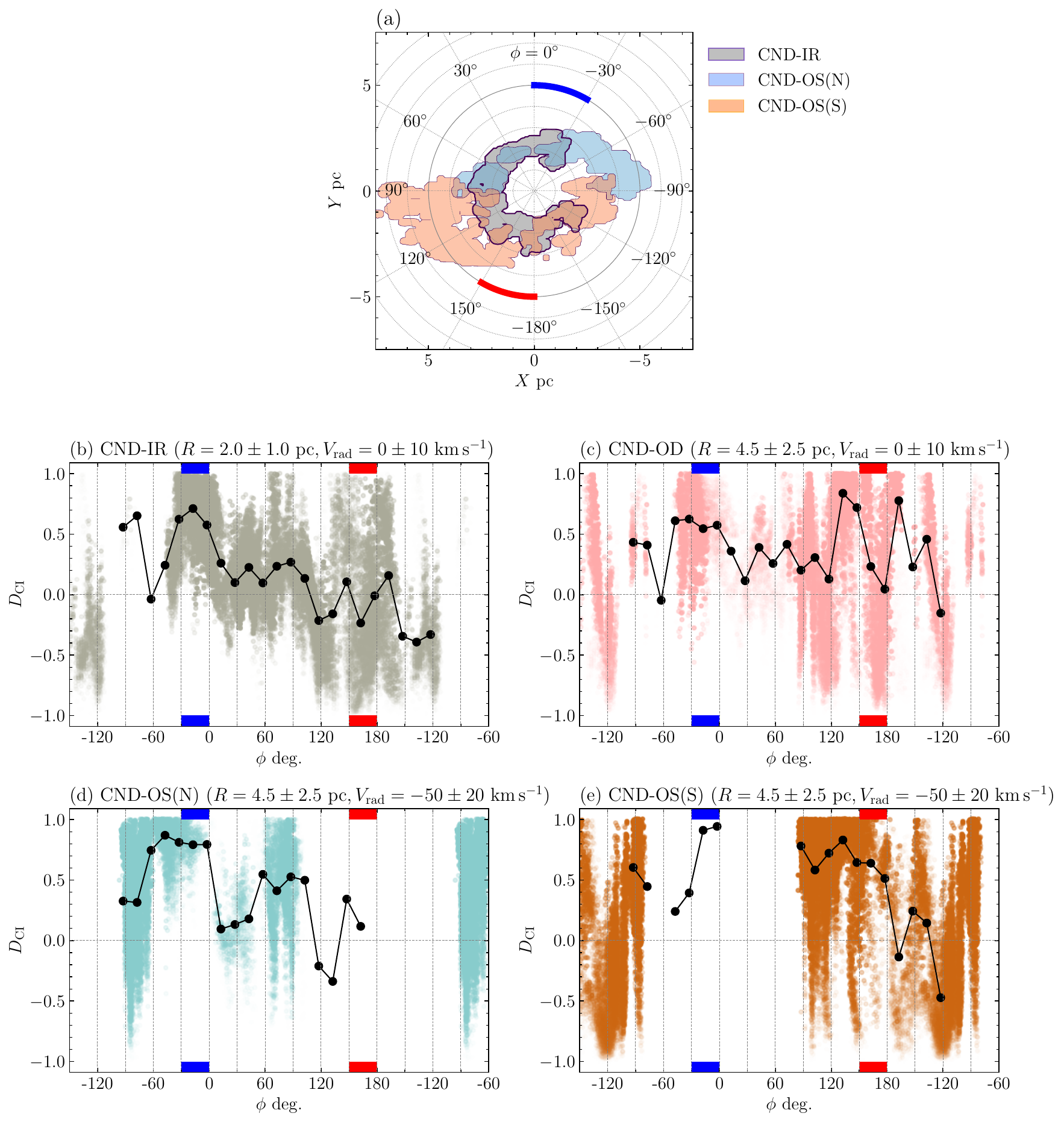}
\caption{\myrevision{Azimuthal  \Dci\ variation.  (a) Analyzed regions on the $X$--$Y$ projection, showing CND-IR, CND-OS(N), and CND-OS(S) after applying $R$--\Vrad\ masks. (b--e) $\phi$--\Dci\ plots for each component.  Black points show \Dci\ values in \ang{15;;}.   The locations and HWHM widths of the $R$--\Vrad\ Gaussian masks are provided in the panel captions.  The $\phi$ ranges of the local \Dci\ humps in CND-IR are highlighted in blue and red in both the $X$--$Y$ and $\phi$--\Dci\ plots.} \label{fig:AzDci}}
\end{figure*}

Figure~\ref{fig:Dci} shows the radial variation of azimuthally averaged \Dci\ in 0.05 pc bins for all CND-associated emission, along with \Dci\ frequency violin plots for the individual components plotted against their average $R$.
The violin plots include the 0-KM and 80-KM anomalies for comparison, although their contributions are excluded from the azimuthal average.
\Dci\ exhibits an overall inwardly decreasing gradient on and outside the CND, with the average \Dci\ decreasing from 0.7 to $-0.2$ as $R$ decreases from 4~\pc\ to 1.7~\pc, while showing a few local humps.
The same radial trend is also observed among the components within this $R$ range, i.e., CND-OS(S)(N), CND-OD, CND-IR, and CND-IS(S), excluding the two anomalies.
\Dci\ remains roughly constant beyond this $R$ range, where the outermost parts of CND-OD and CND-OS are located.
This radial gradient does not persist within $R \sim 1.7$~\pc, inside of which \Dci\ increases toward the center.
This reversal in the \Dci\ gradient is primarily contributed by CND-IS(N).

Figure~\ref{fig:AzDci} shows the azimuthal variation of \Dci\ for CND-IR, CND-OD, and CND-OS(N)(S).
CND-IS is not included in the plot, as its structure deviates significantly from a circular shape.
Overlapping regions among different components are filtered out by applying Gaussian masks in $R$ and \Vrad, with the mask locations and their HWHM widths provided in the panel captions.
As previously discussed in Section~\ref{subsubsection:results:3Dkinematics:IR}, the $\phi$--\Dci\ plot for CND-IR exhibits an overall azimuthal gradient, with \Dci\ decreasing from $\phi=\ang{-30;;}$ to \ang{240;;} (\ang{-120;;} in the plot).
A similar but less clear gradient is seen in the binned plots for CND-OS(N) and CND-OS(S), although their $\phi$ coverages are less uniform and incomplete compared to that of CND-IR.
CND-OD does not show a discernible azimuthal \Dci\ gradient.

The spatial \Dci\ gradient described above is interpreted as a chemical variation along the low-density gas flow: \Dci\ decreases as the gas spirals inward from CND-OS to CND-IR and continues to decline as it rotates along the CND's circular orbit.
Figure~\ref{fig:Dci} shows that this decreasing trend continues in CND-IS(S), which spatially overlaps the downstream end of CND-IR.
This interpretation is consistent with smaller-scale features observed in the azimuthal \Dci\ variation.
\myrevision{
The location where the inflowing gas from CND-OS(N) first contacts CND-IR, indicated by the blue bands in Figure \ref{fig:AzDci}, corresponds to the \Dci\ peak along CND-IR.
We may also identify an increase in high-\Dci\ spaxels near the contacting region between CND-OS(S) and CND-IR, indicated by the red bands in the figure, in the non-binned plot for CND-IR.
This can be reasonably interpreted as a signature that \CI-dominant gas is being supplied to CND-IR by CND-OS.
}

The components that deviate from the global \Dci\ gradient, CND-IS(N) and the two kinematic anomalies, do not align with the above flow model.
The average \Dci\ and frequency profile of CND-IS(N) resemble those of the outermost component, CND-OS, suggesting that CND-IS(N) is not downstream of CND-IS(S) but may represent another flow directly originating from CND-OS(N).
Indeed, the shape of CND-IS(N) (Figure~\ref{fig:regPP}) suggests it bifurcates from the CND-OS(N) mainstream, which merges with CND-IR in projection, and then flows toward the CND interior.
Similarly, the 0-KM and 80-KM anomalies show \Dci\ frequency profiles similar to their adjacent components, CND-IS(N) and (S), respectively, suggesting physical continuity between them.

In addition to providing clues about the low-density gas flow, the spatial variation of \Dci\ offers an important insight into molecular chemistry.
\begin{revision}
Models of photon, X-ray, and cosmic-ray (CR) dominated regions (PDRs, XDRs, and CRDRs, respectively; e.g., \citealt{Bergin1997,Meijerink2005,Bayet2011b}) consistently show that CS is abundant in regions more shielded from these dissociation effects than \Cn\ in the steady state.
However, \Dci\ decreases with decreasing $R$, that is, toward \sgras\ and the central cluster.
This clearly indicates that \Dci, or the \Cn\ abundance, is not a reliable indicator of the UV/X-ray/CR flux, but rather depends on complex factors involving the kinematic processes of molecular gas, in addition to steady-state chemical processes.
\end{revision}

\subsection{Mass Estimate\label{subsection:results:massEstimate}}
This subsection describes mass estimate based on the \CI\ and CS~7--6 luminosity.
We first show that the \CI\ traces cold low-density molecular gas, which was missed in the previous CO spectral line energy distribution (SLED) analysis \citep{Requena-Torres2012}.  
Then we calculate the luminosity-to-mass conversion factors for \CI\ and CS 7--6, upon which the \CI+CS luminosity masses of the individual components are estimated.

\subsubsection{Low-excitation Component\label{subsection:results:excitation}}
Observations indicate that the entire CMZ consists of multiple molecular gas components with different physical properties: a low-excitation component with $\Tkin\sim$20--50~\kelvin\ and $\nHH\sim10^3$~\pcc\ and a high-excitation component typically with $\Tkin\sim$80~\kelvin\ and $\nHH\sim10^{4.1}$~\pcc\ \citep{Tanaka2018b,Tanaka2021}.
The latter high-excitation component associated with the CND is further divided into a moderately dense gas ($\nHH=10^{4.5}$~\pcc, $\Tkin\sim$200~\kelvin) and a warmer, denser gas ($\nHH~\sim10^{5.3}~\pcc$, $\Tkin\sim$300--500~\kelvin) components \citep{Requena-Torres2012}.
\myrevision{Thus, we can interpret the low- and high-excitation components as low- and high-density molecular gas components.}

\begin{figure*}[ttt]
  \begin{center}
    \epsscale{.95}
    \plotone{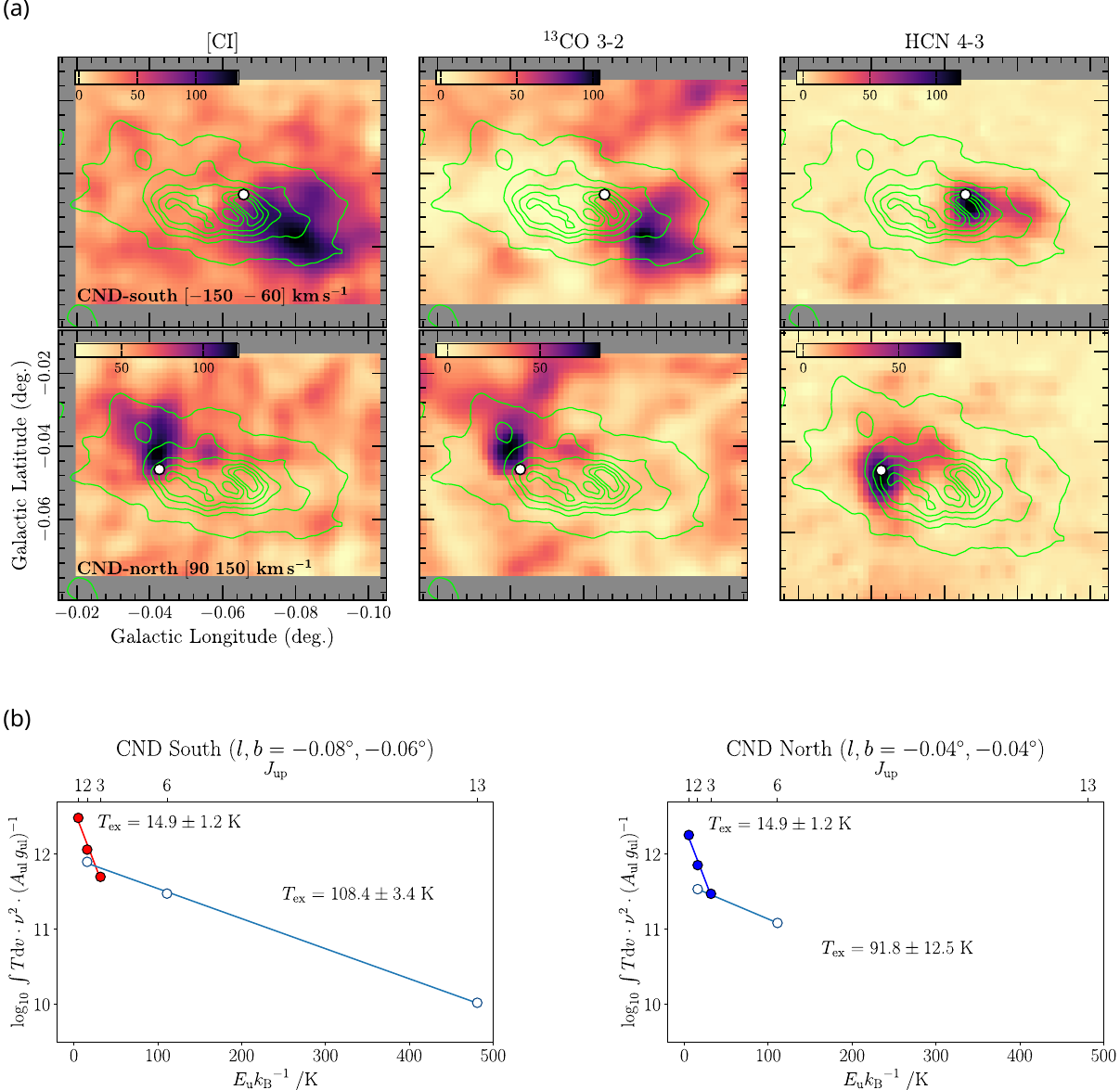}
\end{center}
  \caption{(a) Single-dish images of \CI~\CIa, \COt~3--2 \citep{Tanaka2021}, and HCN~4--3 \citep{Tanaka2018b} integrated intensity in units of $\kelvin\cdot\kmps$ for the CND-south (top row) and CND-north (bottom row). Integration velocity ranges are indicated in the \CI\ panels. Contours show the {\it Hershel} 70~\micron\ image \citep{Molinari2011}, with a contour interval of 10 Jy per $3.2\times3.2~{\rm arcsec}^2$ pixel.  White points mark the positions of the CO~6--5\ peaks \citep{Requena-Torres2012}. (b) \COt\ rotation diagrams for CND-south (left) and CND-north (right).  The filled circles show the \JJx{1}{0}, \JJx{2}{1}, and \JJx{3}{2}\ intensities \citep{Oka1999,Ginsburg2015a, Tanaka2021},  measured toward the \CI\ peaks  within a common beam of $60\arcsec$.  The integration velocity ranges are [$-150$ $-70$]~\kmps\ and [$70$ $100$]~\kmps\ for CND-south and CND-north, respectively.  The open circles represent the \JJx{2}{1}, \JJx{6}{5}, and \JJx{13}{12}\ intensities toward the CO \JJx{6}{5}\ peaks within a common beam of $22.5\arcsec$ \citep{Requena-Torres2012}.  All data points assume a 10\% relative intensity uncertainty. \label{fig:singleDish}}
\end{figure*}

\cite{Tanaka2021} have shown that the \CI\ exhibits a strong spatial correlation with low-$J$ \COt\ transitions over the entire CMZ, while being depleted in the high-excitation component represented by HCN~4--3.
This correlation between the \CI\ and low-$J$ \COt\ transitions persists in smaller spatial scales in the CND.
Figure \ref{fig:singleDish} compares single-dish images of the CND in \CI, \COt~3--2, and HCN~4--3 \citep{Tanaka2018b,Tanaka2021}.
The blue- and red-shifted portions (or CND-south and -north in the nomenclature in \citealt{Requena-Torres2012}) are shown separately in the upper and lower rows, respectively. 
The high-excitation gas represented by HCN~4--3 is almost strictly confined in the CND ring traced by the 70~\micron\ emission \citep{Molinari2011}.
The \CI\ emission peaks are displaced outward from the HCN ring and extend to more than twice the radius of the ring, approximately co-extensive with \COt~3--2\ and the diffuse extended component of the 70~\micron\ emission.
This similarity to \COt~3--2 and displacement from HCN~4--3 confirms that the \CI\ traces the same low-excitation components as low-$J$ \COt\ transitions locally in the CND, as observed for the entire-CMZ scale.

Figure \ref{fig:singleDish}b compares the rotation diagrams of \COt\ at the \CI\ and CO~6--5\ peaks of each of CND-south and -north, using data integrated from literature (\citealt{Oka1999,Ginsburg2015a,Requena-Torres2012,Tanaka2021}; see the figure caption for details).
Two different model lines are overlaid in each panel;  the low-$J$ ($J\leq 3$) diagram toward the \CI~peaks, and the high-$J$ ($J$=2, 6, and 13) diagram toward the CO \JJx{6}{5}\ peaks. 
The low-$J$ diagram is consistently fitted by $\Tex = 15$~$\kelvin$ curves for both CND-south and -north, whereas the high-$J$ diagram is fitted with much higher \Tex\ of $\sim 100$~$\kelvin$.
The latter represents a mixture of the two components in \cite{Requena-Torres2012}.
The low \Tex\ of the \CI-peak indicates that the \CI-emitting gas is the third component of the CND, which was invisible in the CO SLED.
This lowest-excitation component is likely to be the molecular counterpart to the cold dust component with $\Tdust=23.5~\kelvin$ \citep{Etxaluze2011}.

\begin{deluxetable*}{llcccccccc}
\tablecaption{\myrevision{Mass Estimates} \label{table:massEstimates}}
\tablecolumns{10}
\renewcommand{\arraystretch}{1.}
\tablehead{
\multirow{3}{*}{{Region}} & & \multicolumn{2}{c}{{$L_X$ \scriptsize{$\left(\lumunitTab\right)$}}} & \multicolumn{3}{c}{{mass$^{\star}$ \scriptsize{$\left(10^4~\Msun\right)$}}} &  \multirow{3}{*}{\hspace{12pt}\Rlh\hspace{12pt}}  & \multicolumn{2}{c}{{$X_X$ $\left(\XunitTab\right)$}}\\ 
\cmidrule(lr){3-4} \cmidrule(lr){5-7} \cmidrule(lr){9-10}
& & \colhead{\CI} &  \colhead{CS~7--6}  & \colhead{low\tablenotemark{$^\dag$}} & \colhead{high\tablenotemark{$^\ddag$}}& total & & \CI & CS~7--6
}
\startdata 
Inner \ang{;;42.5}         &&      2890\tablenotemark{$^{a}$} &       252\tablenotemark{$^{a}$} &    3.15\tablenotemark{$^{b}$} &  1.56\tablenotemark{$^{b}$} &  4.71 &  2.01 & 10 & 62\\
\hline
\multicolumn{2}{l}{CND and CND-associated components}                      &	   1498 &	    277 &	 1.63 &	 1.71 &	 3.34 &	 0.95 \\
&Inner Ring                  &	  \phn335 &	    \phn81 &	 0.36 &	 0.50 &	 0.87 &	 0.72 \\
&Outer Disk                  &	  \phn168 &	    \phn28 &	 0.18 &	 0.17 &	 0.35 &	 1.07 \\
&Outer Stream                &	  \phn473 &	    \phn66 &	 0.51 &	 0.41 &	 0.92 &	 1.27 \\
&\multicolumn{1}{r}{South}   &   \phn277 &	    \phn42 &	 0.30 &	 0.26 &	 0.56 &	 1.17 \\
&\multicolumn{1}{r}{North}   &   \phn205 &	    \phn19 &	 0.22 &	 0.12 &	 0.34 &	 1.88 \\
&Inner Stream                &	  \phn216 &	    \phn61 &	 0.24 &	 0.38 &	 0.61 &	 0.62 \\
&\multicolumn{1}{r}{South}   &  \phn\phn69 &	    \phn51 &	 0.08 &	 0.31 &	 0.39 &	 0.24 \\
&\multicolumn{1}{r}{North}   &	  \phn147 &	    \phn10 &	 0.16 &	 0.06 &	 0.22 &	 2.53 \\ 
\hline 
\multicolumn{2}{l}{Anomalies}                      & \phn 307 &	   \phn 41 &	 0.33 &	 0.26 &	 0.59 &	 1.30 \\
& 80-\kmps\        &	  \phn108 &	    \phn28 &	 0.12 &	 0.18 &	 0.29 &	 0.67 \\
& 0-\kmps\         &	  \phn199 &	    \phn13 &	 0.22 &	 0.08 &	 0.30 &	 2.53 \\
\hline
\enddata
\tablenotetext{\star}{\myrevision{Mass values include 10--15\%\ systematic and 30\%\ statistical uncertainties.  See text (\S\ref{subsection:results:massEstimate}) for details. } }
\tablenotetext{\dag}{cold dust mass for the inner \ang{;;42.5} region and the \CI\ luminosity mass ($= \Xci\cdot\Lci$) for the others.}
\tablenotetext{\ddag}{hot dust mass for the inner \ang{;;42.5} region and the CS~7--6 luminosity mass  ($= \Xcs\cdot\Lcs$) for the others.}
\tablenotetext{a}{Including both the CND and non-CND components.  See text (\S\ref{subsection:results:massEstimate}). }
\tablenotetext{b}{\cite{Etxaluze2011}}
\end{deluxetable*}

\subsubsection{[CI]+CS luminosity mass}
\cite{Tanaka2021}\ estimated that the low-excitation component of the CND has a high [\Cn]/[CO] abundance ratio of $\sim2$, while the ratio is $\lesssim0.1$ for the high-excitation component.
This allows us to use the \CI\ luminosity as an exclusive mass tracer for the low-excitation component. 
The CS~7--6 emission, on the other hand, can be used to measure the mass of the high-excitation component as its critical density (\ncrit = $10^{6.8}$~\pcc) is nearly $10^4$ times the typical \nHH\ of the low-excitation component.
Hence, we approximate the total molecular \myrevision{hydrogen} mass by the \CI+CS~luminosity mass 
\begin{eqnarray}
\Mmol &=& \Xci\cdot\Lci + \Xcs\cdot\Lcs \label{eqn:massEstimates},     
\end{eqnarray}
where $L_{X}$ and $X_{X}$ denote the luminosity and luminosity-to-mass conversion factor for the transition $X$, respectively. 
The conversion factors are obtained from the \CI\ and CS~7--6 luminosities and the hot and cold dust masses ($1.56\times10^4~\Msun$ and $3.15\times10^4~\Msun$, respectively; \citealt{Etxaluze2011}) within the central \ang{;;42.5}\ circle.
As the dust mass estimate does not distinguish different velocity components, we calculated $X_\mathrm{X}$  based on the total luminosities including both CND and non-CND components.
The mass estimates for the entire CND and individual $R$--\Vrad\ components are tabulated in Table \ref{table:massEstimates} along with \Lci, \Lcs, \Xci, and \Xcs.

We cross-checked the mass estimate by an independent analysis using the single-dish \CI\ data from \cite{Tanaka2021}.
The \CI~luminosity mass integrated over the \vlsr\ ranges of [$-150\ {-40}$]~\kmps\ and [$90\ 150$]~\kmps, where the foreground/background \CI\ emission is negligible, is $3\times10^4~\Msol$, with the assumption that \Tkin = 20~\kelvin, \nHH = $10^3~\pcc$, and [\Cn]/[\molH] = $10^{-4}$. 
In the ALMA data, the \CI\ emission in this velocity range occupies 50\% of the total luminosity of the CND components. 
If the same fraction applies to the single-dish data, the recovered \CI\ luminosity mass for the entire velocity range is  $\sim 6\times10^4~\Msol$.
This is a good agreement with the low-excitation mass in Table \ref{table:massEstimates}, $3.15\times10^4~\Msol$, considering the larger spatial coverage of the single-dish data and uncertainty in the abundance ratios among $\mathrm{H_2}$, \Cn, and dust.

\begin{revision}
Our mass estimates include a systematic uncertainty of 10--15\% inherited from the errors in the dust masses reported by \citet{Etxaluze2011}.
Additionally, a statistical uncertainty among different regions arises from the spatial variation in $X_X$.
In particular, the CS luminosity masses of CND-IR and CND-IS(S) are likely overestimated, as these regions contain bright CS~10--9 emission indicating a higher CS~7--6 excitation than the CND average.
We measured the ratio of CO-based mass \citep{Requena-Torres2012} to CS~7--6 flux to differ by 40\% between the CND-south and CND-north, based on which we roughly estimate the region-to-region variation in \Xcs\ to be $40\%/\sqrt{2}\sim30\%$.
By conservatively assuming a similar variation for \Xci, the statistical uncertainty of the total mass is also expected to be $\sim30\%$.
\end{revision}

Our estimate confirms that the CND and CND-associated clouds contain a significant mass of the low-excitation gas comparable to the high-excitation component mass.
\myrevision{The mass ratio of the low-excitation component to the high-excitation component, \Rlh}, is 0.95 for the entire region average.
It is noteworthy that even CND-IR, which corresponds to the CND's dense-gas ring, has a moderate $\Rlh$ value of 0.69.
The spatial variation of \Rlh\ follows that of \Dci, as evident from the mass estimation process.
In the $R\gtrsim 1.7$~pc region, \Rlh\ decreases with decreasing $R$, from 1.88 in CND-OS(S) to 0.22 in CND-IS(S). However, the innermost component, CND-IS(N), exhibits a high \Rlh\ of $>2$.
The high-excitation gas mass of CND-IR is $0.52\times10^4$\Msun, accounting for only 16\% of the total molecular mass of the CND and CND-associated clouds.
The majority of the molecular mass is associated with the streamer components, whose total mass is $1.53\times10^4$\Msun,  70\% of which is in the low-excitation component.

\section{Discussion\label{section:discussion}}

\subsection{Inner Streamer and the mini-spiral\label{subsection:results:RRLs}}
The CND and its associated features are considered the primary source of mass supply to the ionized gas comprising the mini-spiral.
Several feeding channels have been proposed based on dense-gas observations.
One such channel is the interface between the southwestern rim of the CND and the mini-spiral WA, where molecular and ionized emissions show strong \PPV\ correlation along nearly the entire extent of the WA \citep[e.g.,][]{Guesten1987, Christopher2005, Tsuboi2018}.
Another is anomaly-B \citep{Tsuboi2018}, a dense-gas filament spiraling from outside the CND to the eastern arm (EA) of the mini-spiral.
Additionally, \citet{Tsuboi2020} detected molecular emission along the mini-spiral NA, suggesting a physical connection with molecular gas outside the central cavity.

\begin{figure*}
\epsscale{1.15}
\plotone{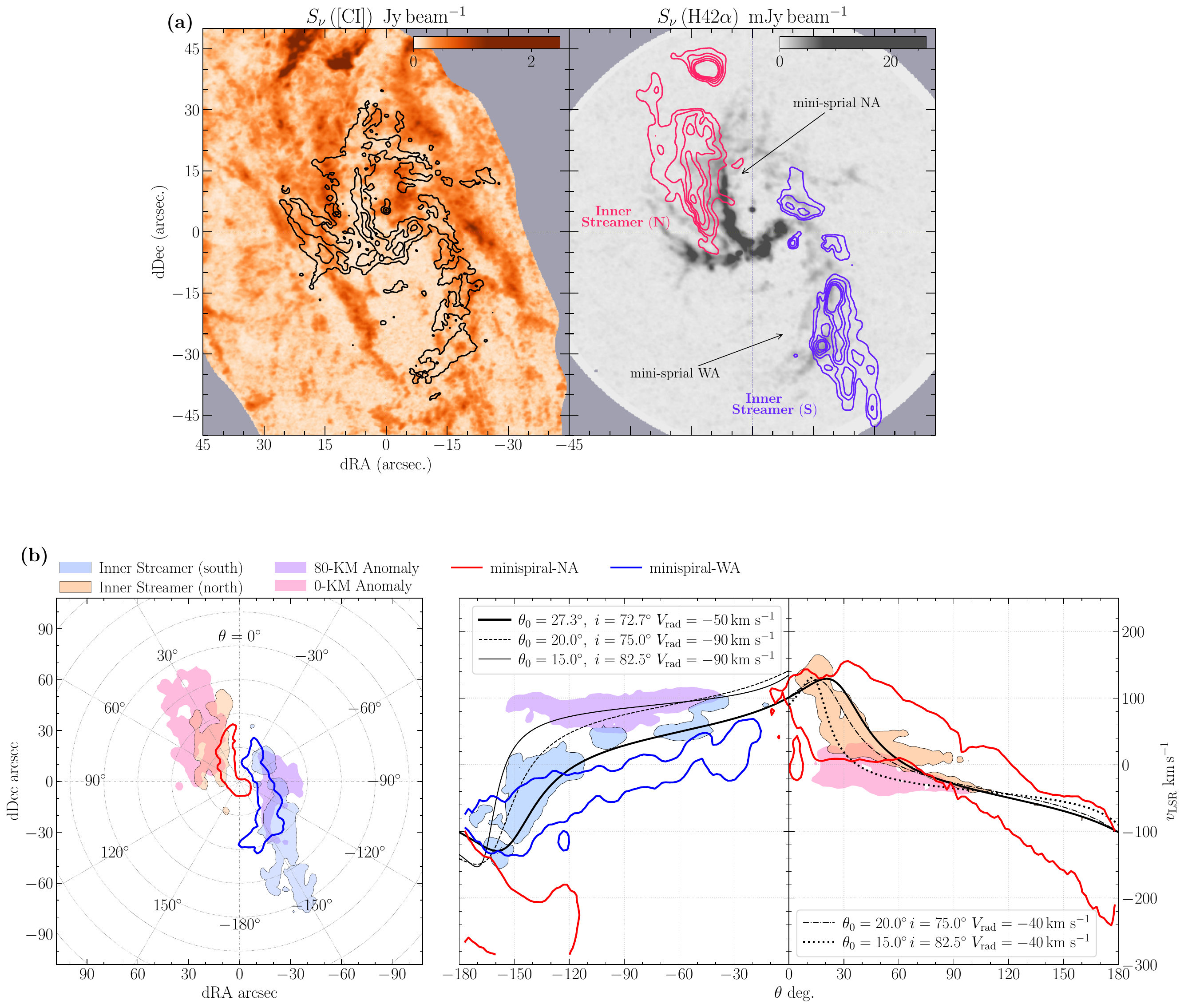}
\caption{\myrevision{(a) Peak flux densities of \CI\ (left) and H42$\alpha$ (right; ALMA archival data 2013.1.00857.S, 2016.1.00940.S, and 2022.1.01219.S), measured in 10-\kmps\ and 25-\kmps\ velocity bins, respectively.  On the \CI\ image,  H42$\alpha$ contours are overlaid at 0.05, 0.2, and 0.5 times the peak value, 28\ m\jkpb\ (with a synthesized beam of $\ang{;;1.13}\times\ang{;;0.94}$ and \ang{-83.7}\ position angle).  On the H42$\alpha$ image, \CI\ contours of CND-OS(N) and (S) are overlaid at 0.1, 0.2, 0.3, 0.4, and 0.5 times their respective peak values (2.3 and 1.0 \jkpb).  (b) \PPV\ distributions of 0-KM and 80-KM anomalies, CND-IS, and the mini-spiral NA and WA on the RA--Dec projection (left) and the $\theta$--\vlsr\ phase plot (right). Overlaid curves represent model orbits described in \S\ref{subsubsection:results:3Dkinematics:RVdiagram} with various combinations of $i$, $\theta$, and $\theta_0$, and fixed \mstar\ ($0.8\times10^6$~\Msun) and $R$ (2~\pc).}
\label{fig:miniSpiral}}
\end{figure*}

In our data, the inflow toward the mini-spiral is traced by CND-IS, the \CI\ component located inside the CND with large \Vrad.
We compare the \PPV\ distributions of \CI\ and H42$\alpha$ using peak intensity maps and the $\theta$--\vlsr\ phase plot shown in Figures~\ref{fig:miniSpiral}a and b, respectively.
The H42$\alpha$ data were obtained from the ALMA science archive. 
Our analysis indicates that the CND side of the known CND--WA interface exhibits infalling motion.
CND-IS(S) extends along WA and is systematically offset westward by approximately $\ang{;;5}$.
\CI\ and H42$\alpha$ also show a strong correlation in \vlsr\ variation, with a systematic velocity offset of approximately $-10$ to $-20$~\kmps\ from WA to CND-IS(S).
Since the velocity of CND-IS(S) lies between those of WA and the CND's stable rotational orbit, this spatial and velocity offset is naturally interpreted as a transition from dense molecular gas to ionized gas.

We identify a similar \PPV\ correlation between the NA and CND-IS(N).
The NA comprises a vertical emission ridge and diffuse emission extending eastward from the ridge.
Unlike the WA, no spatial association with the CND ring has previously been reported for the NA in dense-gas observations.
However, Figure~\ref{fig:miniSpiral}a shows that the NA ridge outlines the eastern edge of CND-IS(N) over an angular span of approximately $\ang{;;30}$.
The diffuse H42$\alpha$ component appears to fill the gap between the NA and the CND, and its spatial extent closely matches that of CND-IS(N).
In the $\theta$--\vlsr\ phase diagram (Figure~\ref{fig:miniSpiral}b), the NA and CND-IS(N) exhibit correlated \vlsr\ variations from $\theta = \ang{30;;}$ to $\ang{120;;}$, with H42$\alpha$ having \vlsr\ values similar to or up to $\sim+30$~\kmps\ higher than those of \CI.
As with the WA and CND-IS(S), this correlation suggests that the ionized gas forming the NA originates from the inner edge of CND-IS(N).

Thus, our $R$--\Vrad\ diagram successfully distinguishes molecular gas components undergoing rapid infall from the CND from the purely rotating ring,
confirming the previously identified connection between the mini-spiral WA and the infalling \CI\ feature CND-IS(S),
and newly revealing that the mini-spiral NA likewise extends from CND-IS(N).
We failed to confirm the association between anomaly-B and the EA that was previously reported in dense-gas imaging, although their transition region likely falls within CND-OS(N) in our data.
This is presumably because the transition region consists of moderately dense gas, which is bright in CS~2--1 emission \citep{Tsuboi2018}, whereas its \CI\ emission is expected to be fainter than that of the main component of CND-OS(N).
Given that this region exhibits a complex spatial and velocity structure with multiple overlapping components, EA-associated \CI\ emission may remain undetectable if it lacks sufficient brightness.

\subsection{Molecular Gas Flows To and From the CND \label{subsection:discussion:flow} }

Our ALMA data reveal a large amount of low-density ($\nHH\sim10^3$~\pcc) molecular gas both inside and outside the CND ring.
Most of the low-density features, except for CND-OD, exhibit fast inward radial velocities of $\sim$40--70~\kmps, forming a global inflow that connects the CND exterior to the mini-spiral. The masses of the inflowing components outside and inside the CND are $0.91\times10^3$~\Msun\ and $0.62\times10^3$~\Msun, respectively, each comparable to the CND-IR mass of $0.88\times10^3$~\Msun. The total inflowing mass is $1.5\times10^4$~\Msun, approximately 100 times the previous estimates based on dense-gas tracers \citep{Hsieh2017}. Given the limited FoV of our observations, the actual inflowing mass could be even larger. In the wider-field \CI\ image \citep{Tanaka2021}, the emission within the ALMA FoV accounts for only 30~\% of the total luminosity in the contamination-free \vlsr\ ranges (i.e., [$-150$ $-40$]~\kmps\ and [$90$ $150$]~\kmps; see \S\ref{subsection:results:excitation}). Including this out-of-FoV \CI\ emission would increase the inflowing mass by a factor of 2--3, even conservatively omitting the mass of the 0-KM and 80-KM anomalies ($0.59\times10^3$~\Msun). Hence, the overall structure of the low- and high-excitation molecular gas associated with the CND is better described as a globally infalling system toward the mini-spiral, with the CND ring comprising only a relatively small fraction in mass.

\begin{figure*}
\epsscale{1}
\plotone{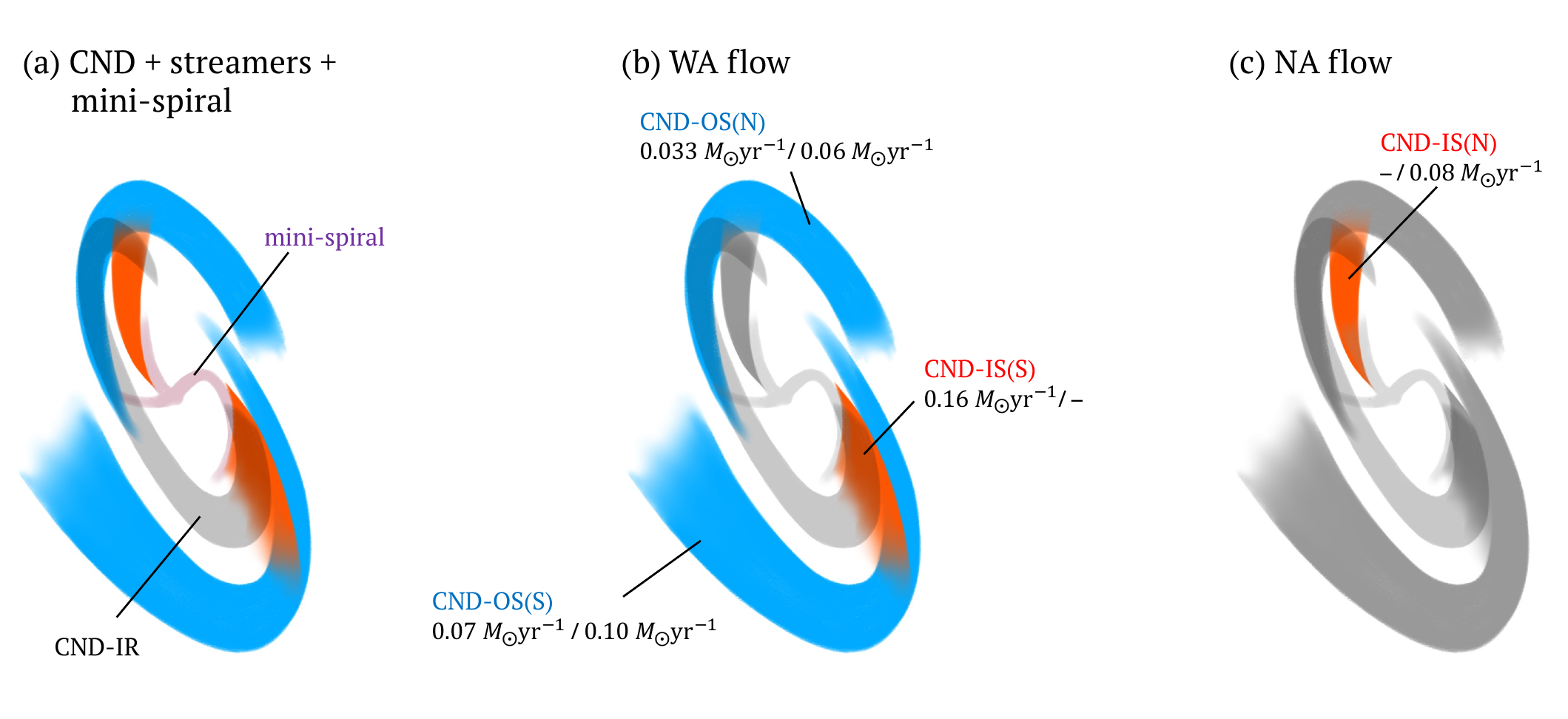}
\caption{
(a) Schematic drawing of the CND, CND-associated inflows, and the mini-spiral on the RA--Dec projection.
(b) and (c) Same as (a), but highlighting the WA and NA flows, respectively.
The numbers under the flow label indicate the inflow rates estimated using Equations~\ref{eqn:massFluxAz} and \ref{eqn:massFluxRad}, represented as $\dot{M}_\mathrm{az}$/$\dot{M}_\mathrm{rad}$\myrevision{, each including $\gtrsim30$~\%\ uncertainty inherited from the uncertainties in the mass estimate.}
\label{fig:flows}}
\end{figure*}

As we have seen in \S\ref{subsection:results:CItoCSratio} and \S\ref{subsection:results:massEstimate}, \Dci\ (or the \CI\ to CS~7--6 intensity ratio) primarily depends on the mass fraction of the low-excitation to high-excitation components, rather than on local environmental parameters such as UV, X-ray, or CR fluxes.
Assuming that the radial and azimuthal gradients of \Dci\ arise from the supply of high-\Dci\ gas from outer radii, we identify the major paths of molecular gas streams associated with the CND, which we schematically show in Figure \ref{fig:flows}.

The flows represented by the CND-OS, IR, and IS can be sorted into two main inflowing paths.
The first path consists of spiral flows originating from outer radii ($R\gtrsim 5~\pc$) that merge with CND-IR and eventually reach the mini-spiral WA.   We refer to this path, which transits the CND ring, as the ``WA flow.''
CND-OS(N) and (S), CND-IS(S), and CND-IR are included in the WA flow.
As their emission extends beyond the FoV of our observations, the starting points of these outer flows are not identifiable in our data.
Previous wider-field observations suggest that the upstreams of these CND-associated streamers connect to the 50-\kmps\ and 20-\kmps\ clouds \citep{Hsieh2017,Takekawa2017}.
The second path is the ``NA flow,'' which directly feeds the mini-spiral without merging with the flow along CND-IR.
CND-IS(N) represents this path, which originates near the location where CND-OS first contacts CND-IR and immediately flows toward the interior of CND-IR.
In addition, we may identify the third pathway, the ``EA flow'' connecting the Anomaly-B and the mini-spiral EA, which was discovered by \cite{Tsuboi2018} with dense-gas observation, although we failed to clearly detect this channel in the \CI\ data.

Figure \ref{fig:flows} shows the mass inflow rate ($\dot{M}$) of the individual streams estimated in two ways.
The first method uses the mass flux in the azimuthal direction: 
\begin{eqnarray} 
\dot{M}_\mathrm{az} &\sim& \frac{\Delta M}{\Delta \phi}\Omega_{2\pc}, \label{eqn:massFluxAz} 
\end{eqnarray} 
where $\Omega_{2\pc} = 2\pi\times10^{-5}~\mathrm{rad}\,\yr^{-1}$ is the circular frequency at $R=2$~\pc, $\Delta\phi$ is the azimuthal span of the overlapping region with CND-IR, and $\Delta M$ is the mass associated with the overlapping region.
Equation \ref{eqn:massFluxAz} applies to streams that have a significant circular portion overlapping CND-IR.
The second method is based on the mass flux in the radial direction: \begin{eqnarray} \dot{M}_\mathrm{rad} &\sim& \frac{M}{\Delta R}\left|{\Vrad}\right|, \label{eqn:massFluxRad} \end{eqnarray} where $M$ and $\Delta R$ are the total mass and radial extent of the infalling component, respectively.
This estimate is applicable for streams with a spiral or radially linear shape, whose $\Delta R$ is sufficiently larger than their intrinsic transverse widths.
We apply Equation~\ref{eqn:massFluxAz} to CND-IS(S), Equation~\ref{eqn:massFluxRad} to CND-IS(N), and both Equations~\ref{eqn:massFluxAz} and \ref{eqn:massFluxRad} to CND-OS(N) and CND-OS(S).
$\dot{M}_\mathrm{az}$ and $\dot{M}_\mathrm{rad}$ are approximately consistent for CND-OS, although the latter tends to give slightly higher inflow rates.

The estimated inflow rates indicate that CND-IR is rapidly gaining low-density molecular gas through the WA flow.
With a total inflow rate of $\sim0.1~\Msun\,\yr^{-1}$ from CND-OS(N) and (S), CND-IR will accumulate molecular gas comparable to its current mass ($0.88\times10^4$~\Msun) within one orbital period at $R$=2~\pc\ ($\sim10^5$~\yr).
The mass loss rate through CND-IS(S) is $0.1~\Msun\,\yr^{-1}$, which approximately balances the inflow rate.
\begin{revision}
This implies that the dwelling time of gas in CND-IR is comparable to the CND's orbital period, supporting the CND's transient nature.
\end{revision}
The inflow rate of the NA flow through CND-OS(N) is estimated to be $0.08~\Msun\,\yr^{-1}$, comparable to the inflow rates of the WA flow.
When combined with the EA flow, which was not identified with our analysis, the flows bypassing the CND form a major channel of molecular gas transport within the CND system, alongside the WA flow transiting the CND.

CND-OD does not exhibit significant inflow motion, but its high \Dci\ value and spatial proximity to CND-OS suggest a physical association with the WA flow.
CND-OD may be interpreted as a low-density gas disk, supplied by CND-OS and settled into rotating orbits at $R > 2$~\pc\ before reaching the CND orbit.
Whether the rotating disk of CND-OD is a stable structure remains uncertain.
However, its morphology deviates significantly from an axisymmetric disk, with a large fraction of the \CI\ emission concentrated near the junctions of CND-OS and CND-IR ($\phi \sim \ang{-30;;}$ and \ang{150;;}).
This non-uniform azimuthal distribution suggests that the disk may be short-lived.

\begin{revision}
The high mass flux of $\sim 0.1~\Msun\yr^{-1}$ is approximately a factor of a few to an order of magnitude higher than the theoretically calculated rate of steady mass loss of the CND \citep{Vollmer2002,Tress2020}.
This implies that the CND system has not yet reached a dynamically settled state after the most recent GMC capture event, where fast inward motion could be facilitated by collisions between different streams.
%
Two scenarios can be considered for the origin of this situation.
One possibility is that the CND-IR and the streamers were formed in a single GMC capture event. The observed spiral pattern along the WA flow resembles that found in numerical models of single GMC capture \citep{Sanders1998,Bonnell2008,Mapelli2016}.
Our results show that the CND-IR and CND-OS share a common orbital plane, supporting the idea that they inherited the angular momentum of the same progenitor GMC.
In this scenario, the CND-IR can be regarded as a transient structure emerging within the WA flow, which eventually connects to the mini-spiral.
This scenario may also explain the NW gap of the CND-IR, whose endpoints approximately coincide with the contact point between CND-OS(N) and CND-IR (i.e., where gas inflow into the CND-IR begins) and the downstream end of CND-IS(S) (i.e., where gas outflow toward the mini-spiral WA ends).
An alternative scenario is that the WA flow consists of a combination of a pre-existing dense-gas ring and a newly infalling low-density stream, similar to the configuration assumed in the model by \citet{Alig2013}.
This multi-infall scenario naturally explains the coexistence of two types of features with contrasting physical and kinematical properties, namely, a steadily rotating dense-gas ring and low-density infalling streamers, by assuming two progenitor GMCs.
The observed \Dci\ gradient along the WA flow can then be understood as gradual mixing of the pre-existing CS-dominant gas with newly infalling \CI-dominant gas.

\end{revision}

\subsection{0-KM and 80-KM~anomalies: non-CND clouds or nearly edge-on streamers\label{subsection:results:anomalies}}

The 0-KM and 80-KM anomalies lack azimuthal velocity gradient over spatial scales of 2--4~\pc, contradicting the CND's rotation.
A straightforward interpretation is that these anomalies are foreground or background clouds similar to the western filament, despite their projected morphologies resembling CND-associated emission.
However, a drawback of this interpretation is that their systemic velocities lie outside the typical \vlsr\ range of the known overlapping GMCs (20--60~\kmps).

Alternatively, the anomalies may be interpreted as CND-associated features moving on a nearly edge-on orbital plane.
Figure~\ref{fig:miniSpiral}b shows model curves overlaid on the $\theta$--\vlsr\ phase plot, based on the same kinematic framework described in \S\ref{subsubsection:results:3Dkinematics:RVdiagram}, but with varying values of $i$, $\theta_0$, and \Vrad.
The model parameters are detailed in the figure caption.
As $i$ approaches $\ang{90;;}$, the line-of-sight component of circular motion diminishes at $\phi$ away from the inclination axis, making it less discernible.
We find that orbits with high inclination with $i = \ang{75;;}$--$\ang{82.5;;}$ and $\theta_0 = \ang{15;;}$--$\ang{20;;}$ reproduce the observed velocity patterns of both anomalies when assuming appropriate values of \Vrad: $-90$~\kmps\ for the 0-KM anomaly and $-40$~\kmps\ for the 80-KM anomaly.

If the edge-on cloud interpretation is correct, the two anomalies may be regarded as part of the NA and WA flows.
Not only do they exhibit similar \PP\ distributions to CND-IS(N) and CND-IS(S), but they also show similar \Dci\ frequency profiles to their respective neighboring CND-IS components, suggesting their physical continuity, as we have seen in \S\ref{subsection:results:CItoCSratio}.
Since they move along planes that are significantly inclined relative to the main CND orbit, the anomalies are unlikely to be the immediate upstream sources of CND-IS(N) and (S).
Nonetheless, it remains plausible that interactions between the anomalies and CND-OS or CND-IR have triggered rapid infall motions toward the mini-spiral.

However, we caution that a successful fit with high-$i$ models in itself does not confirm their validity, as high-$i$ orbits can reproduce almost any flat-velocity feature with a suitable choice of $\theta_0$ and \Vrad.
This ambiguity weakens the argument for associating the anomalies with the CND, especially given the large \vlsr\ differences of $\sim$50--100~\kmps.
Hence, we leave the relation between the anomalies and the CND unconcluded in this paper.

\subsection{\CI-bright Clumps: Dense Clumps forming in Colliding Flows? \label{subsection:results:CBCs}\label{subsection:discussion:CBCs}}

\begin{deluxetable*}{l@{\hspace{0.5\tabcolsep}}l @{\hspace{8\tabcolsep}} r@{\,}l @{\hspace{8\tabcolsep}} r @{\hspace{0\tabcolsep}}l l}

\tablecolumns{6}
\tablecaption{\CI-bright Clumps  \label{tab:CIclumps}}
\tablehead{
\colhead{} & & \multicolumn{2}{l}{CBC-A} & \multicolumn{2}{l}{CBC-B} 
}
\startdata 
Peak $\mathrm{H_2}$ column density& ($\NHH$)$^{\dagger}$    & \hfill 3.8 & $\times 10^{23}\,\mathrm{cm}^{-2}$ \hfill & 2.4 & $\times 10^{23}\,\mathrm{cm}^{-2}$ \\
\CI+CS\ luminosity mass& (\Mmol)$^{\dagger}$       & \hfill 4.4    & {$\times 10^2\,\Msun$} \hfill      & 6.0 & {$\times 10^2\,\Msun$}      \\
Clump radius& ($\rclump$) & \hfill 0.12    &\ \pc \hfill & 0.18    &\ \pc \\
Distance from \sgras& ($R$)       & \hfill 2.0 &\ \pc \hfill & 2.2 &\ \pc \\
Mean $\mathrm{H_2}$ volume density& ($\nHH$)$^{\dagger}$   & \hfill 5.5 & {$\times 10^5\,\pcc$} \hfill & 2.3 &  {$\times 10^5\,\pcc$} \\
Tidal critical density& ($\nroche$) & \hfill 6.1 & {$\times 10^6\,\pcc$} \hfill & 5.7 & {$\times 10^6\,\pcc$}
\enddata
\tablenotetext{\dagger}
Includes 10--15\% systematic and 30\% statistical uncertainties inherited from the mass estimates; see \S\ref{subsection:results:massEstimate} and Table~\ref{table:massEstimates} (footnote).
\end{deluxetable*}

\begin{figure*}
\epsscale{1.}
\plotone{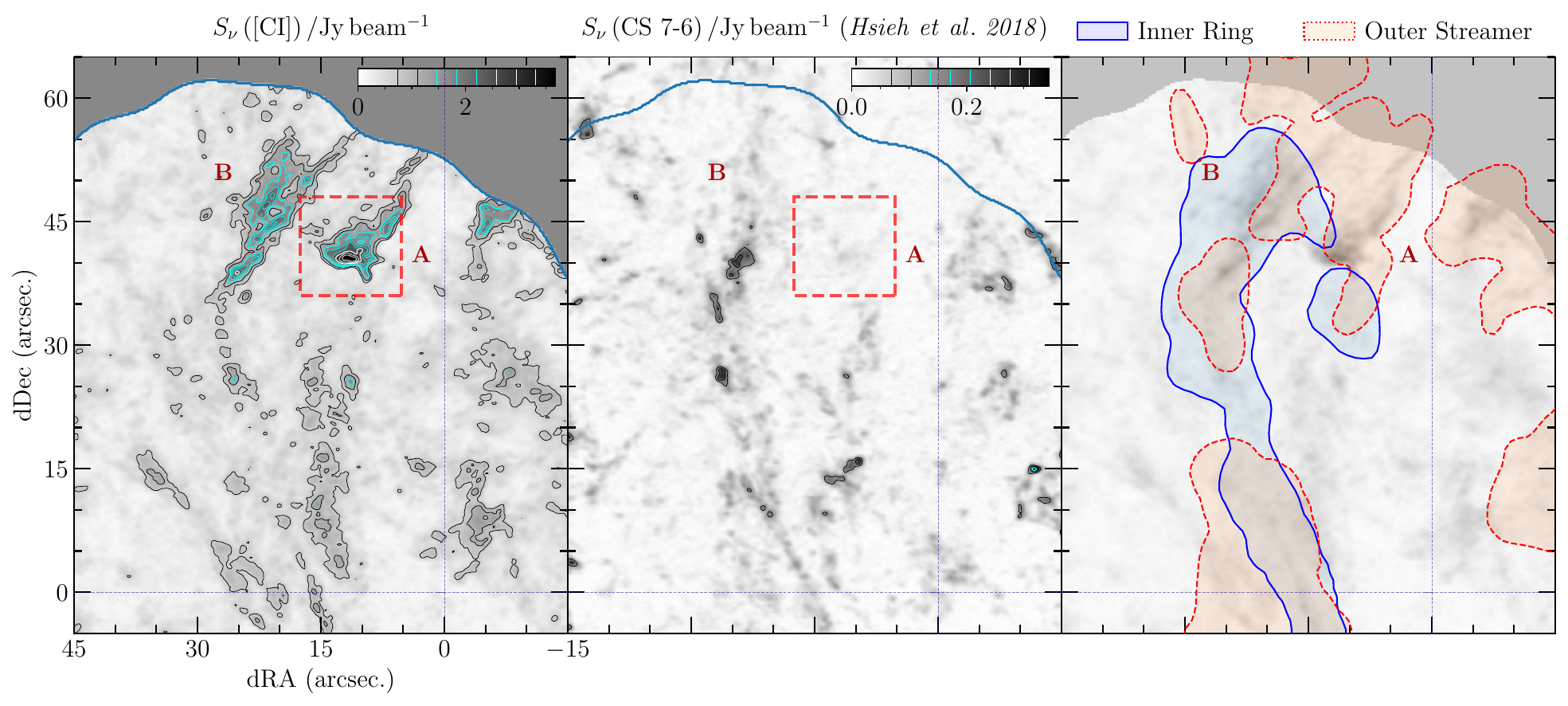}
\caption{CBCs-A and -B in the \CI\ (left) and CS~7--6 (middle) peak intensity maps, with contours drawn at (0.2, 0.3, ..., 0.9) times the maximum flux densities.  The right panel schematically shows CND-IR and CND-OS on the \CI\ intensity image. \label{fig:CIclump}} 
\end{figure*}

In \S\ref{subsection:results:images}, we identified two CBCs exhibiting remarkably bright \CI\ emission without CS counterparts.
Figure~\ref{fig:CIclump} shows their close-up views in the \CI\ and CS~7--6 images.
A schematic representation of CND-IR and CND-OS is overlaid, indicating that the CBCs are located in the overlapping region of these two components.
Their radius (\rclump), peak column density (\NHH), \nHH, and \Mmol\ are summarized in Table \ref{tab:CIclumps}, along with the tidal critical density (\nroche).
The \nHH\ values are estimated assuming a line-of-sight depth of $2 \rclump$, where \rclump\ is defined as $\rclump\equiv\sqrt{\Mmol/\left(2.8\pi\mu_\mathrm{H}\NHH\right)}$ with $\mu_\mathrm{H}$ being proton mass.

Both CBCs have high \NHH\ values of a few $10^{23}~\psc$, comparable to the self-gravitating region of the Brick cloud \citep{Rathborne2014a}; indeed, the relatively globular shape of CBC-A appears ostensibly unaffected by tidal shear.
Their densities, \nHH\ $\sim$ a few $\times10^5$~\pcc, are two orders of magnitude higher than those of the typical \CI-emitting gas estimated in \S\ref{subsection:results:excitation}, and are more consistent with the high-excitation component.
However, these values are still a factor of 10--20 below \nroche, indicating that the CBCs cannot maintain their compact \NHH\ peaks over a timescale of $\ttidal \sim \left(R,\mathrm{d}\Omega_R/\mathrm{d}R\right)^{-1} \sim 10^4$~yr, where $\Omega_R = \sqrt{G M(R)/R^3}$ is the circular frequency.
This limits their lifetime to $\lesssim10^4$~yr, which is shorter than both the circular orbital period $2\pi \Omega_R^{-1} \sim 10^5$~yr and the inflow timescale for CND-OS, $\sim 2~\pc / \Vrad = 4\times10^4$~yr.
Therefore, the CBCs are not stable clumps, but likely transient structures that formed near their current locations and will dissipate on short timescales.
Indeed, the elongation of CBC-B along the direction of the circular orbit suggests that it is already on the verge of tidal disruption.

The combination of high \nHH\ and high \Dci\ in the CBCs contradicts the two-component model adopted so far, which posits a \CI-emitting low-density component and a CS-emitting high-density component.
This suggests that the CBC's exceptionally high \Dci\ value of $\sim1$ cannot be explained by the primary parameters in static UV/X-ray/CR dissociation models, namely, \nHH\ and the strength of the dissociative effects.
However, the tidally subcritical nature of the CBCs offers an explanation in terms of non-stationary chemistry.
The chemical timescale for the [\Cn]/[CO] abundance ratio to reach equilibrium is $\sim10^4$~\yr\ in gas with $\nHH\sim10^5$~\pcc\ \citep{Bergin1997}.
Since this timescale is comparable to $\ttidal$ of $\sim10^4$~\yr, tidally unstable clumps can retain their high-\Dci\ state inherited from the low-density phase for a substantial portion of their lifetime.
The CBCs may therefore represent clumps in transition from \CI-dominant, accreting gas to the CS-dominant gas that constitutes the CND ring.

\begin{revision}
This discovery of the CBCs has implications for the origin of the young stellar population in the central parsec.
The \textit{in-situ} formation scenario describes efficient compression of molecular gas in turbulent accretion flows over a wide range of $R$ \citep{Paumard2004,Bonnell2008,Hobbs2009,Alig2013,Mapelli2016}.
Figure~\ref{fig:CIclump} shows that the CBCs lie near the boundary between CND-IR and CND-OS, where shock compression from colliding molecular flows is expected.
This indicates that shock-driven formation of dense clumps is possible in the presence of harsh tidal field, providing observational support for the \textit{in-situ} formation scenario.
Although the detected CBCs themselves are unlikely to contribute to the young stellar population, they could evolve into star-forming clumps if they were located at $R\sim6~\pc$, where their \nHH\ would be comparable to \nroche.
Since single-dish observations have shown that the \CI\ emission extends out to $\sim10~\pc$, we may speculate that young, star-forming clumps exist that remained undetected in previous observations.
On the other hand, we have seen that the dense-gas structure of CND-IR is sufficiently captured using only the CS image, as \Dci\ in CND-IR is generally $\lesssim 0$ except in the region overlapping with CND-OS(N).
The evaluation of the tidal stability and self-gravitational instability of the CND ring is therefore essentially unchanged from previous results, in which the contribution of the \CI-derived mass was omitted.

\end{revision}

\section{Summary\label{section:summary}}

We present ALMA images of \CI~$^3P_1$--$^3P_0$ and CS~10--9 emission toward the central  $6.6\times4.2~\mathrm{pc}^2$ including the CND and its associated features. 
Based on these data, we investigate the distribution and kinematics of low-excitation molecular gas that were not well captured in earlier observations relying primarily on high-density tracers.
The main results are summarized below:

\begin{itemize} 
\item \CI\ emission is widespread both inside and outside the CND ring, with a larger weight toward the northern half of the FoV. It exhibits only weak correlation with the dense-gas structures represented by CS~7--6. \CI-bright features lacking CS counterparts, such as the NE extension, SE Arc, and the \CI-bright clumps (CBCs), are detected (\S\ref{subsection:results:images}).

\item A substantial portion of the \CI\ emission does not follow the purely rotating motion of the CND (\S\ref{subsection:results:nonCNDfeatures}). By interpreting deviations from circular motion as radial velocities (\Vrad), the \CI\ spaxels are projected into orbital-plane coordinates ($R$, $\phi$). Besides two anomalies likely located outside the CND’s orbital plane, four kinematic components are identified in the $R$--\Vrad\ diagram by SMM: two purely rotating (CND-IR and CND-OD) and two infalling (CND-OS and CND-IS).

\item CND-IR corresponds to the CND ring, rotating at 125~\kmps\ on the orbital plane with PA=$\bestAlp$ and $i=\bestI$ (\S\ref{subsubsection:results:3Dkinematics:IR}). Its outer envelope is accompanied by another purely rotating component, CND-OD, consisting of low-density gas sparsely distributed over $R\sim$3--6~\pc\ (\S\ref{subsubsection:results:3Dkinematics:OD}).

\item CND-OS comprises a pair of spiral streamers infalling from $R\sim$7~\pc\ to 1.5~\pc\ with an average \Vrad\ of $\sim -39.3$~\kmps. The northern and southern arms (CND-OS(N) and CND-OS(S)) approximately correspond to the known features Western streamer and Anomaly B, respectively (\S\ref{subsubsection:results:3Dkinematics:OS}).

\item CND-IS contains inflowing material in the CND interior with $\Vrad=68.5~\kmps$. Its southern component (CND-IS(S)) spatially overlaps with the SW portion of CND-IR but is distinguished by its fast inward motion. The northern component (CND-IS(N)) extends inward from the NE extension of the CND to within $\sim$0.5~\pc\ of \sgras\ (\S\ref{subsubsection:results:3Dkinematics:IS}).
The \PPV\ correlation with H42$\alpha$ emission suggests that CND-IS(N) and (S) are feeding paths of molecular gas to the mini-spiral NA and WA, respectively (\S\ref{subsection:results:RRLs}).

\item The normalized \CI\ to CS~7--6 intensity difference (\Dci) decreases radially from $R=4$~\pc\ to 1.7~\pc\ and azimuthally along the CND-IR orbit. This overall \Dci\ gradient contradicts simple UV/X-ray/CR dissociation models but aligns with the low-density gas flow inferred from the \Vrad\ distribution. The anomalously high \Dci\ in CND-IS(N), deviating from the global radial gradient, suggests an alternative flow that bypasses CND-IR and directly reaches the interior (\S\ref{subsection:results:CItoCSratio}).

\item From a rotation diagram using low-$J$ $^{13}$CO data from the literature, we estimate \Tkin\ and \nHH\ of the \CI-emitting gas to be $\sim$30~\kelvin\ and $\sim10^3$~\pcc. The \CI\ luminosity mass of the low-excitation gas in the CND is at least comparable to the CS~7--6 luminosity mass of the high-excitation component. The streamer components (CND-OS and CND-IS) contain a total mass of $1.53\times10^4$~\Msun, about 1.7 times that of CND-IR within the ALMA FoV (\S\ref{subsection:results:massEstimate}).

\item Two velocity components at \vlsr\ = 0~\kmps\ and 80~\kmps, named the 0-KM and 80-KM anomalies, cannot be fitted with the rotating/streaming model on the same orbital plane as the CND (\S\ref{subsubsection:results:3Dkinematics:0and80kmComp}). They can be modeled as infalling clouds on nearly edge-on planes, though a non-CND origin cannot be ruled out (\S\ref{subsection:discussion:flow},\ref{subsection:results:anomalies}).

\item The streamers inside and outside the CND trace two distinct paths. The ``WA flow'' spirals in from CND-OS, passes through the CND's circular orbit, and ultimately feeds the mini-spiral WA via CND-IS(S). The ``NA flow'' bypasses CND-IR and directly feeds the mini-spiral NA. The inflow rate is nearly constant along the WA flow ($\sim$0.1--0.16~\Msun\,yr$^{-1}$), and the dwelling time within CND-IR is comparable to its orbital period, suggesting a transient nature of CND-IR (\S\ref{subsection:discussion:flow}).

\item CBC-A and -B are compact ($r \sim 0.1$--$0.2$~\pc), \CI-bright clumps without CS counterparts, newly discovered near the contact points between CND-OS and CND-IR. Despite their apparently undisturbed morphology, their \nHH\ ($\sim 2$--$6\times10^5$~\pcc) is an order of magnitude below their tidal critical density, implying they formed recently ($\lesssim 10^5$~\yr) near their present locations. The combination of high \nHH\ and high \Dci\ ($\sim 1$) suggests they are in a chemically non-equilibrium state, representing a transitional phase from infalling, \CI-dominant gas to dense molecular gas (\S\ref{subsection:discussion:CBCs}).

\end{itemize}

\begin{acknowledgments}
\myrevision{
The authors are grateful to the anonymous reviewer for many helpful comments and suggestions that improved the paper.
}
This paper makes use of the following ALMA data: ADS/JAO.ALMA\#2015.1.01040.S, \#2013.1.00857.S, \#2016.1.00940.S, and \#2022.1.01219.S.
ALMA is a partnership of ESO (representing its member states), NSF (USA) and NINS (Japan), together with NRC (Canada), MOST and ASIAA (Taiwan), and KASI (Republic of Korea), in cooperation with the Republic of Chile. The Joint ALMA Observatory is operated by ESO, AUI/NRAO and NAOJ.
The authors acknowledge the use of the AI tool ChatGPT (versions 3.5 and 4.0) exclusively for language editing of the manuscript. 
\end{acknowledgments}

\begin{appendix}
\newcommand\vvec{\boldsymbol{v}}
\newcommand\Fvec{\boldsymbol{F}}
\section{Fitting of the orbit parameters \label{appendix:fitting}}

This appendix describes the Bayesian analysis to solve the coordinate conversion described by Equations \ref{eqn:proj1}--\ref{eqn:vrad}.
The posterior function is constructed as follows:
\begin{eqnarray}
	\Prob{\vartheta|\mathcal{V}} & = & - \frac{1}{\sigma_v^2}\sum\limits_m w_m^2 \left[\vlsrm - \left(\Vradm\cdot\sin\phi_m + \Vrot\left(R_m\right)\cdot\cos\phi_m\right)\sin i\right]^2 + p\cdot f_1\left(\vartheta\right) + f_2\left(\vartheta\right) \label{eqn:posterior} \\
    f_1\left(\vartheta\right) &=& -\sum\limits_{\left<m,n\right>} w_m w_n \left(\Vradm - \Vradn\right)^2 \label{eqn:smoothnessPrior} \\
    f_2\left(\vartheta\right) &=& -\frac{1}{2}\sum\limits_{m} 
    \begin{cases}
    \left(\frac{\vlsrm}{v_\mathrm{max}}\right)^2 & \left(\Vradm \geq 0\right) \\ 
    \left(\frac{\vlsrm}{v_\mathrm{min}}\right)^2 & \left(\Vradm < 0\right)
    \end{cases},  \label{eqn:rangePrior}
\end{eqnarray}
where $\mathcal{V}=\left(\Vradm\right)$ and $\vartheta = \left(\vlsrm, R_m\right)$, with subscript $m$ indexing each \PPV\ spaxel.  
For the weight $w_m$, we use the \CI\ intensity at the $m$th spaxel. 
The parameter $i$ is the orbital inclination, and $\Vrot\left(R\right)$ is the circular velocity at $R$.
The summation in Equation \ref{eqn:smoothnessPrior} is taken over all adjacent spaxel pairs $\left<m, n \right>$ without duplication.
The first term in $\Prob{\vartheta|\mathcal{V}}$ represents logarithm of the likelihood function $\mathcal{L}\left(\mathcal{V}|\mathcal{\vartheta}\right)$ assuming a normal error distribution for $\mathcal{V}$ with variance $\sigma_v^2/2$.
We adopt $\sigma=10~\kmps$, which is approximately a typical velocity dispersion of the \CI\ emission. 
The second and third terms, $f_1\left(\vartheta\right)$ and $f_2\left(\vartheta\right)$, are prior functions regularizing the \PPV\ variation of \Vrad.
The first prior $f_1\left(\vartheta\right)$ constrains the variance of \Vrad\ among adjacent spaxels to be minimized in the \textit{a priori} distribution.
This enables us to infer an optimal value for \Vrad\ at $\phi = \ang{0;;}, \ang{180;;}$, where $\mathcal{L}\left(\mathcal{V}|\vartheta\right)$ does not depend on \Vrad.
The strength of $f_1\left(\vartheta\right)$ is controlled by the hyperparameter $p$. 
When $p=0$, the optimal $\mathcal{V}$ is essentially the same as that provided by Equation \ref{eqn:vrad}, while $p=\infty$ enforces a constant $\Vrad$ across the entire \PPV\ volume.
The second prior $f_2\left(\vartheta\right)$ imposes a weak constraint on the dynamic range of $\vlsr$.   We use limits $v_{\mathrm{min,max}} = -100~\kmps,\ +10~\kmps$ for the present analysis.
The full $\Prob{\vartheta|\mathcal{V}}$ distribution is calculated by employing the Markov Chain Monte Carlo method.
The median of the marginalized posterior is adopted as the representative value of each \Vradm. 

\begin{figure*}
\epsscale{1.05}
\plotone{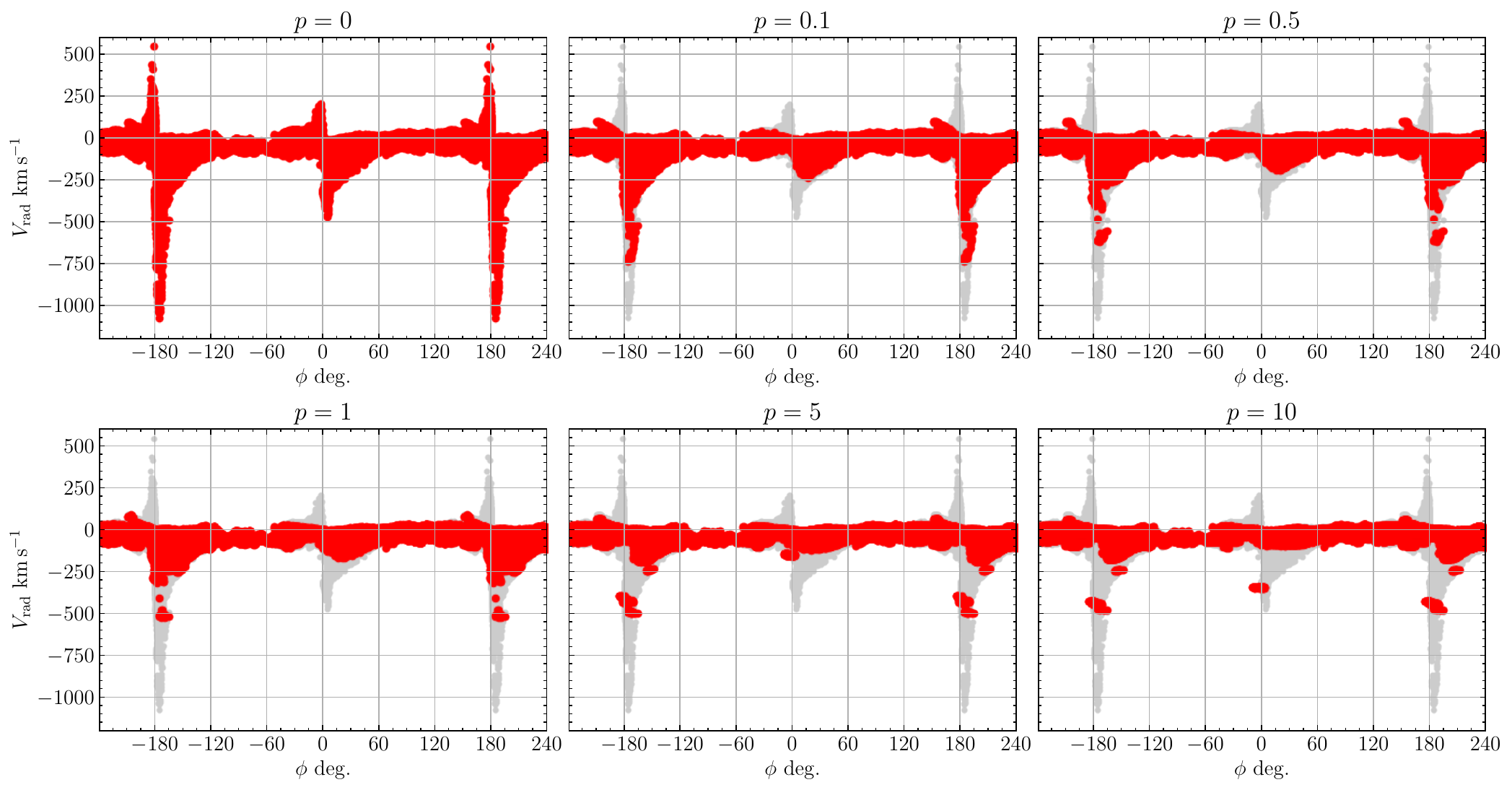}
\caption{ $\phi$--\Vrad\ plots with $p$ = 0, 0.1, 0.5, 1, 5, and 10.  The results with $p=0$ are overlaid in gray for the results with $p>0$.   \label{fig:orbitParamsDiffP}}
\end{figure*}

The hyperparameter $p$ can be arbitrarily chosen  so that the divergence at $\Vrad$ at $\phi = \ang{0;;}$ and $\ang{180;;}$ is sufficiently mitigated.
In general, larger $p$ values yield smoother $\Vrad$ distribution, while the observational information is more respected by solutions with $p$ closer to 0.  
Figure \ref{fig:orbitParamsDiffP} compares $\phi$--$\Vradm$ plots with different $p$ ranging from 0 to 10, with fixed orbital parameters of $\theta_0 = \bestAlp$ and $i = \bestI$. 
The result with $p=0$ exhibits large discontinuity across $\phi = \ang{0;;}$ and \ang{180;;} as the consequence of the $\sim\sin(\phi)^{-1}$ dependence of $\Vrad$ in Equation \ref{eqn:vrad}.
This discontinuity diminishes as $p$ increases; however, horizontal features with large negative \Vrad, likely artifacts resulting from an overly strong smoothness prior, start to appear for $p \gtrsim 1$.  
We adopt $p$ = 1 for the present analysis, which achieves a satisfactorily smooth \Vrad\ distribution without introducing pronounced artificial features.

\myrevision{
As we discussed in \S\ref{subsubsection:results:3Dkinematics:RVdiagram}, we ignore the vertical extent, $\Delta z$, from the orbital plane in translating the plane-of-the-sky coordinate into the $\phi$--$R$ coordinate on the orbital plane.  
The error in \Vrad\ is expected to have $\sim\sin2\phi$ dependence.   However, Figure \ref{fig:orbitParamsDiffP} does not exhibit significant increase in \Vrad\ dispersion around $\phi = \pm\ang{45;;}$ and $\pm\ang{135;;}$ even with $p=0$.  This confirms that the effect of $\Delta z$ on the $\Vrot$ estimate is most comparable to the intrinsic velocity width of the \CI\ emission, and hence is safely ignored in our analysis. 
}

\section{\myrevision{Student-$t$ Mixture Model} \label{appendix:SMM}}
\myrevision{
The Student-$t$ mixture model decomposes data points in the $p$-dimensional space ($p$=2 in the present analysis), which we express by vector $\myvector{x}_i$ ($i$=1,2,..., $N_p$, where $N_p$ is the number of the data points), into a given number ($K$) of clusters by assuming that the probability density distribution $\rho\left(\myvector{x}\right)$ is expressed as a weighted superposition of multivariate Student-$t$ kernels;
}

\begin{eqnarray}
\rho\left(\myvector{x}\right) & \propto & \pi_j\sum_{j=1,2,..., K} t_\nu\left(\myvector{x}|\myvector{\mu}_j, \Sigma_j, \nu\right),\\
 t_\nu\left(\myvector{x}|\myvector{\mu}, \Sigma, \nu\right) & = &\left|\Sigma\right|^{-\frac{1}{2}} \left\{1 + \left(\myvector{x}-\myvector{\mu}\right)^T \Sigma \left(\myvector{x}-\myvector{\mu}\right)\right\}^{-\frac{\nu+p}{2}}. \label{eqn:studentT}
\end{eqnarray}
Equation \ref{eqn:studentT} omits the normalization constants here for the purpose of conciseness.  The optimal weights $\pi_j$, location vectors $\myvector{\nu}_j$, and scale matrices $\Sigma_j$ are iteratively obtained to maximize the likelihood $\mathcal{L}\left(\myvector{x}_1, \myvector{x}_2, ..., \myvector{x}_{N_p}|\pi_1, \pi_2, ..., \pi_K, \myvector{\mu}_1, \myvector{\mu}_2, ..., \myvector{\mu}_K, \Sigma_1, \Sigma_2, ..., \Sigma_K \,\right) = \prod_{i} \rho\left(\myvector{x}_i\right)$, for which we adopted the EM algorithm \citep{Dempster1977}.  We confirmed that the deconvolution results are not dependent on the initial parameter for our problem by performing 100 analysis runs starting from randomly chosen $\pi_j$, $\myvector{\mu}_j$, and $\Sigma_j$. 
The probability for the $i$th data point to belong to the $j$th component is represented by the responsibility $\gamma_{ij}$ given by 
\begin{eqnarray}
\gamma_{ij} & \equiv & \frac{\pi_j t_\nu\left(\myvector{x}_i|\myvector{\mu}_j, \Sigma_j, \nu\right)}{\sum_k \pi_k t_\nu\left(\myvector{x}_i|\myvector{\mu}_k, \Sigma_k, \nu\right)},
\end{eqnarray}
namely, the relative contribution of $j$th kernel to the probability of the data point having the given $\myvector{x}_i$.  
\color{black}
\end{appendix}

\bibliographystyle{apj}
\bibliography{mendeley,local}

\newcommand{\noop}[1]{}
\begin{thebibliography}{}
\expandafter\ifx\csname natexlab\endcsname\relax\def\natexlab#1{#1}\fi

\bibitem[{Alig {et~al.}(2013)Alig, Schartmann, Burkert, \& Dolag}]{Alig2013}
Alig, C., Schartmann, M., Burkert, A., \& Dolag, K. 2013, \apj, 771, 119

\bibitem[{Ballone {et~al.}(2019)Ballone, Mapelli, \& Trani}]{Ballone2019}
Ballone, A., Mapelli, M., \& Trani, A.~A. 2019, \mnras, 488, 5802

\bibitem[{Barna {et~al.}(2025)Barna, W{\"{u}}nsch, Palous, Morris,
  Ehlerov{\'{a}}, \& Vermot}]{Barna2025}
Barna, B., W{\"{u}}nsch, R., Palous, J., {et~al.} 2025, \aap, 695, A92

\bibitem[{Bayet {et~al.}(2011)Bayet, Williams, Hartquist, \& Viti}]{Bayet2011b}
Bayet, E., Williams, D.~A., Hartquist, T.~W., \& Viti, S. 2011, \mnras, 414,
  1583

\bibitem[{Bergin {et~al.}(1997)Bergin, Goldsmith, Snell, \&
  Langer}]{Bergin1997}
Bergin, E.~A., Goldsmith, P.~F., Snell, R.~L., \& Langer, W.~D. 1997, \apj,
  482, 285

\bibitem[{Blank {et~al.}(2016)Blank, Morris, Frank, Carroll-Nellenback, \&
  Duschl}]{Blank2016}
Blank, M., Morris, M.~R., Frank, A., Carroll-Nellenback, J.~J., \& Duschl,
  W.~J. 2016, \mnras, 459, 1721

\bibitem[{Bonnell \& Rice(2008)}]{Bonnell2008}
Bonnell, I.~A., \& Rice, W.~K. 2008, Science, 321, 1060

\bibitem[{{CASA Team} {et~al.}(2022){CASA Team}, {Bean}, {Bhatnagar}, {Castro},
  {Donovan Meyer}, {Emonts}, {Garcia}, {Garwood}, {Golap}, {Gonzalez Villalba},
  {Harris}, {Hayashi}, {Hoskins}, {Hsieh}, {Jagannathan}, {Kawasaki},
  {Keimpema}, {Kettenis}, {Lopez}, {Marvil}, {Masters}, {McNichols},
  {Mehringer}, {Miel}, {Moellenbrock}, {Montesino}, {Nakazato}, {Ott}, {Petry},
  {Pokorny}, {Raba}, {Rau}, {Schiebel}, {Schweighart}, {Sekhar}, {Shimada},
  {Small}, {Steeb}, {Sugimoto}, {Suoranta}, {Tsutsumi}, {van Bemmel},
  {Verkouter}, {Wells}, {Xiong}, {Szomoru}, {Griffith}, {Glendenning}, \&
  {Kern}}]{CASA2022}
{CASA Team}, {Bean}, B., {Bhatnagar}, S., {et~al.} 2022, \pasp, 134, 114501

\bibitem[{Christopher {et~al.}(2005)Christopher, Scoville, Stolovy, \&
  Yun}]{Christopher2005}
Christopher, M.~H., Scoville, N.~Z., Stolovy, S.~R., \& Yun, M.~S. 2005, \apj,
  622, 346

\bibitem[{Coil \& Ho(1999)}]{Coil1999}
Coil, A.~L., \& Ho, P. T.~P. 1999, \apj, 513, 752

\bibitem[{Coil \& Ho(2000)}]{Coil2000}
---. 2000, \apj, 533, 245

\bibitem[{{Dempster} {et~al.}(1977){Dempster}, {Laird}, \&
  {Rubin}}]{Dempster1977}
{Dempster}, A.~P., {Laird}, N.~M., \& {Rubin}, D.~B. 1977, Journal of the Royal
  Statistical Society, Series B (Methodological), 39, 1

\bibitem[{Etxaluze {et~al.}(2011)Etxaluze, Smith, Tolls, Stark, \&
  Gonz{\'{a}}lez-Alfonso}]{Etxaluze2011}
Etxaluze, M., Smith, H.~A., Tolls, V., Stark, A.~A., \& Gonz{\'{a}}lez-Alfonso,
  E. 2011, \aj, 142, 134

\bibitem[{Garc{\'{i}}a {et~al.}(2016)Garc{\'{i}}a, Simon, Stutzki,
  G{\"{u}}sten, Requena-Torres, \& Higgins}]{Garcia2016}
Garc{\'{i}}a, P., Simon, R., Stutzki, J., {et~al.} 2016, \aap, 588, 131

\bibitem[{Genzel {et~al.}(1985)Genzel, Watson, Crawford, \&
  Townes}]{Genzel1985}
Genzel, R., Watson, D.~M., Crawford, M.~K., \& Townes, C.~H. 1985, \apj, 297,
  766

\bibitem[{Ginsburg {et~al.}(2015)Ginsburg, Walsh, Henkel, Jones, Cunningham,
  Kauffmann, Pillai, Mills, Ott, Kruijssen, Menten, Battersby, Rathborne,
  Contreras, Longmore, Walker, Dawson, \& Lopez}]{Ginsburg2015a}
Ginsburg, A., Walsh, A., Henkel, C., {et~al.} 2015, \aap, 584, L7

\bibitem[{Goicoechea {et~al.}(2018)Goicoechea, Pety, Chapillon, Cernicharo,
  Gerin, Herrera, Requena-Torres, \& Santa-maria}]{Goicoechea2018a}
Goicoechea, J.~R., Pety, J., Chapillon, E., {et~al.} 2018, \aap, 618, 35

\bibitem[{{Gravity Collaboration; Abuter} {et~al.}(2019){Gravity Collaboration;
  Abuter}, Amorim, Berger, Bonnet, Brandner, Zeeuw, Dexter, Duvert, Eckart,
  Eisenhauer, Garc{\'{i}}a, Gao, Gendron, Genzel, Gerhard, Gillessen, Habibi,
  Haubois, Henning, Hippler, Horrobin, Jocou, Kervella, Lacour, Ott, Paumard,
  Perraut, Perrin, Pfuhl, Rabien, Coira, Rousset, Scheithauer, Sternberg,
  Straub, Straubmeier, Sturm, Tacconi, Vincent, Waisberg, Widmann, Wieprecht,
  Wiezorrek, Woillez, \& Yazici}]{GravityCollaboration2019}
{Gravity Collaboration; Abuter}, R., Amorim, A., Berger, J.~P., {et~al.} 2019,
  \aap, 625, L10

\bibitem[{Guesten {et~al.}(1987)Guesten, Genzel, Wright, Jaffe, Stutzki, \&
  Harris}]{Guesten1987}
Guesten, R., Genzel, R., Wright, M. C.~H., {et~al.} 1987, \apj, 318, 124

\bibitem[{Herrnstein \& Ho(2005)}]{Herrnstein2005}
Herrnstein, R.~M., \& Ho, P. T.~P. 2005, \apj, 620, 287

\bibitem[{Heywood {et~al.}(2019)Heywood, Camilo, Cotton, Abbott, Adam, \&
  Aldera}]{Heywood2019}
Heywood, I., Camilo, F., Cotton, W.~D., {et~al.} 2019, \nature, 573, 235

\bibitem[{Hobbs \& Nayakshin(2009)}]{Hobbs2009}
Hobbs, A., \& Nayakshin, S. 2009, \mnras, 394, 191

\bibitem[{Hsieh {et~al.}(2015)Hsieh, Ho, \& Hwang}]{Hsieh2015}
Hsieh, P.-Y., Ho, P. T.~P., \& Hwang, C.-Y. 2015, \apj, 811, 142

\bibitem[{Hsieh {et~al.}(2017)Hsieh, Koch, Ho, Kim, Tang, Wang, Yen, \&
  Hwang}]{Hsieh2017}
Hsieh, P.-Y., Koch, P.~M., Ho, P. T.~P., {et~al.} 2017, \apj, 847, 3

\bibitem[{Hsieh {et~al.}(2018)Hsieh, Koch, Kim, Ho, Tang, \& Wang}]{Hsieh2018}
Hsieh, P.-Y., Koch, P.~M., Kim, W.-T., {et~al.} 2018, \apj, 862, 150

\bibitem[{Hsieh {et~al.}(2019)Hsieh, Koch, Kim, Ho, Yen, Harada, \&
  Tang}]{Hsieh2019}
Hsieh, P.-Y., Koch, P.~M., Kim, W.~T., {et~al.} 2019, \apjl, 885, L20

\bibitem[{Hsieh {et~al.}(2021)Hsieh, Koch, Kim, Mart{\'{i}}n, Yen, Carpenter,
  Harada, Turner, Ho, Tang, \& Beck}]{Hsieh2021a}
Hsieh, P.-Y., Koch, P.~M., Kim, W.-T., {et~al.} 2021, \apj, 913, 94

\bibitem[{Irons {et~al.}(2012)Irons, Lacy, \& Richter}]{Irons2012}
Irons, W.~T., Lacy, J.~H., \& Richter, M.~J. 2012, \apj, 755, 90

\bibitem[{Jackson {et~al.}(1993)Jackson, Geis, Genzel, Harris, Madden,
  Poglitsch, Stacey, \& Townes}]{Jackson1993}
Jackson, J.~M., Geis, N., Genzel, R., {et~al.} 1993, \apj, 402, 173

\bibitem[{Liu {et~al.}(2013)Liu, Ho, Wright, Su, Hsieh, Sun, Kim, \&
  Minh}]{Liu2013}
Liu, H.~B., Ho, P. T.~P., Wright, M. C.~H., {et~al.} 2013, \apj, 770, 44

\bibitem[{Liu {et~al.}(2012)Liu, Hsieh, Ho, Su, Wright, Sun, \& Minh}]{Liu2012}
Liu, H.~B., Hsieh, P.-Y., Ho, P. T.~P., {et~al.} 2012, \apj, 756, 195

\bibitem[{Lo \& Claussen(1983)}]{Lo1983}
Lo, K.~Y., \& Claussen, M.~J. 1983, Nature, 306, 647

\bibitem[{Mapelli \& Trani(2016)}]{Mapelli2016}
Mapelli, M., \& Trani, A.~A. 2016, \aap, 585, 161

\bibitem[{Marr {et~al.}(1993)Marr, Wright, \& Backer}]{Marr1993}
Marr, J.~M., Wright, M. C.~H., \& Backer, D.~C. 1993, \apj, 411, 667

\bibitem[{Mart{\'{i}}n {et~al.}(2012)Mart{\'{i}}n, Martin-Pintado,
  Montero-Casta{\~{n}}o, Ho, \& Blundell}]{Martin2012}
Mart{\'{i}}n, S., Martin-Pintado, J., Montero-Casta{\~{n}}o, M., Ho, P. T.~P.,
  \& Blundell, R. 2012, \aap, 539, 29

\bibitem[{Meijerink \& Spaans(2005)}]{Meijerink2005}
Meijerink, R., \& Spaans, M. 2005, \aap, 409, 397

\bibitem[{Molinari {et~al.}(2011)Molinari, Bally, Noriega-Crespo,
  Compi{\`{e}}gne, Bernard, Paradis, Martin, Testi, Barlow, Moore, Plume,
  Swinyard, Zavagno, Calzoletti, {Di Giorgio}, Elia, Faustini, Natoli,
  Pestalozzi, Pezzuto, Piacentini, Polenta, Polychroni, Schisano, Traficante,
  Veneziani, Battersby, Burton, Carey, Fukui, Li, Lord, Morgan, Motte,
  Schuller, Stringfellow, Tan, Thompson, Ward-Thompson, White, \&
  Umana}]{Molinari2011}
Molinari, S., Bally, J., Noriega-Crespo, a., {et~al.} 2011, \apj, 735, L33

\bibitem[{Montero-Casta{\~{n}}o {et~al.}(2009)Montero-Casta{\~{n}}o,
  Herrnstein, \& Ho}]{Montero-Castano2009}
Montero-Casta{\~{n}}o, M., Herrnstein, R.~M., \& Ho, P. T.~P. 2009, \apj, 695,
  1477

\bibitem[{Moser {et~al.}(2017)Moser, S{\'{a}}nchez-Monge, Eckart,
  Requena-Torres, Garc{\'{i}}a-Marin, Kunneriath, Zensus, Britzen, Sabha,
  Shahzamanian, Borkar, \& Fischer}]{Moser2016}
Moser, L., S{\'{a}}nchez-Monge, {\'{A}}., Eckart, A., {et~al.} 2017, \aap, 603,
  68

\bibitem[{Namekata {et~al.}(2009)Namekata, Habe, Matsui, \&
  Saitoh}]{Namekata2009}
Namekata, D., Habe, A., Matsui, H., \& Saitoh, T.~R. 2009, \apj, 691, 1525

\bibitem[{Oka {et~al.}(1999)Oka, White, Hasegawa, Sato, Tsuboi, \&
  Miyazaki}]{Oka1999}
Oka, T., White, G.~J., Hasegawa, T., {et~al.} 1999, \apj, 515, 249

\bibitem[{Paumard {et~al.}(2004)Paumard, Maillard, \& Morris}]{Paumard2004}
Paumard, T., Maillard, J.~P., \& Morris, M. 2004, \aap, 426, 81

\bibitem[{Paumard {et~al.}(2006)Paumard, Genzel, Martins, Nayakshin,
  Beloborodov, Levin, Trippe, Eisenhauer, Ott, Gillessen, Abuter, Cuadra,
  Alexander, \& Sternberg}]{Paumard2006}
Paumard, T., Genzel, R., Martins, F., {et~al.} 2006, \apj, 643, 1011

\bibitem[{Rathborne {et~al.}(2014)Rathborne, Longmore, Jackson, Foster,
  Contreras, Garay, Testi, Alves, Bally, Bastian, Kruijssen, \&
  Bressert}]{Rathborne2014a}
Rathborne, J.~M., Longmore, S.~N., Jackson, J.~M., {et~al.} 2014, \apj, 786,
  140

\bibitem[{Requena-Torres {et~al.}(2012)Requena-Torres, G{\"{u}}sten, Wei{\ss},
  Harris, Mart{\'{i}}n-Pintado, Stutzki, Klein, Heyminck, \&
  Risacher}]{Requena-Torres2012}
Requena-Torres, M.~A., G{\"{u}}sten, R., Wei{\ss}, a., {et~al.} 2012, \aap,
  542, L21

\bibitem[{Sanders(1998)}]{Sanders1998}
Sanders, R.~H. 1998, \mnras, 294, 35

\bibitem[{Sch{\"{o}}del {et~al.}(2018)Sch{\"{o}}del, Gallego-Cano, Dong,
  Nogueras-Lara, Gallego-Calvente, Amaro-Seoane, \& Baumgardt}]{Schodel2018}
Sch{\"{o}}del, R., Gallego-Cano, E., Dong, H., {et~al.} 2018, \aap, 609, A27

\bibitem[{Serabyn \& Lacy(1985)}]{Serabyn1985}
Serabyn, E., \& Lacy, J.~H. 1985, \apj, 293, 445

\bibitem[{{Sofue} {et~al.}(2025){Sofue}, {Oka}, {Longmore}, {Walker},
  {Ginsburg}, {Henshaw}, {Bally}, {Barnes}, {Battersby}, {Colzi}, {Ho},
  {Jimenez-Serra}, {Kruijssen}, {Mills}, {Petkova}, {Sormani}, {Wallace},
  {Armijos-Abendano}, {Dutkowska}, {Enokiya}, {Fukui}, {Garcia}, {Guzman},
  {Henkel}, {Hsieh}, {Hu}, {Immer}, {Jeff}, {Klessen}, {Kohno}, {Krumholz},
  {Lipman}, {Martin}, {Morris}, {Nogueras-Lara}, {Nonhebel}, {Ott}, {Pineda},
  {Requena-Torres}, {Rivilla}, {Riquelme-Vasquez}, {Sanchez-Monge},
  {Santa-Maria}, {Smith}, {Tanvir}, {Tolls}, \& {Wang}}]{Sofue2025}
{Sofue}, Y., {Oka}, T., {Longmore}, S.~N., {et~al.} 2025, \pasj, 77, 687

\bibitem[{Solanki {et~al.}(2023)Solanki, Ressler, Murchikova, Stone, \&
  Morris}]{Solanki2023}
Solanki, S., Ressler, S.~M., Murchikova, L., Stone, J.~M., \& Morris, M.~R.
  2023, \apj, 953, 22

\bibitem[{Stark {et~al.}(1991)Stark, Gerhard, Binney, \& Bally}]{Stark1991}
Stark, A.~A., Gerhard, O.~E., Binney, J., \& Bally, J. 1991, \mnras, 248, 14

\bibitem[{Takekawa {et~al.}(2017)Takekawa, Oka, \& Tanaka}]{Takekawa2017}
Takekawa, S., Oka, T., \& Tanaka, K. 2017, \apj, 834, 121

\bibitem[{Tanaka {et~al.}(2021)Tanaka, Nagai, \& Kamegai}]{Tanaka2021}
Tanaka, K., Nagai, M., \& Kamegai, K. 2021, \apj, 915, 79

\bibitem[{Tanaka {et~al.}(2018)Tanaka, Nagai, Kamegai, Iino, \&
  Sakai}]{Tanaka2018b}
Tanaka, K., Nagai, M., Kamegai, K., Iino, T., \& Sakai, T. 2018, \apjs, 236, 40

\bibitem[{Tanaka {et~al.}(2011)Tanaka, Oka, Matsumura, Nagai, \&
  Kamegai}]{Tanaka2011}
Tanaka, K., Oka, T., Matsumura, S., Nagai, M., \& Kamegai, K. 2011, \apjl, 743,
  L39

\bibitem[{Tress {et~al.}(2020)Tress, Sormani, Glover, Klessen, Battersby,
  Clark, {Perry Hatchfield}, \& Smith}]{Tress2020}
Tress, R.~G., Sormani, M.~C., Glover, S.~C., {et~al.} 2020, \mnras, 499, 4455

\bibitem[{Tsuboi {et~al.}(2020)Tsuboi, Kitamura, Tsutsumi, Miyawaki, Miyoshi,
  \& Miyazaki}]{Tsuboi2020}
Tsuboi, M., Kitamura, Y., Tsutsumi, T., {et~al.} 2020, \pasj, 72, L5

\bibitem[{Tsuboi {et~al.}(2018)Tsuboi, Kitamura, Uehara, Miyawaki, Miyoshi, \&
  Miyazaki}]{Tsuboi2018}
Tsuboi, M., Kitamura, Y., Uehara, K., {et~al.} 2018, \pasj, 70, 85

\bibitem[{Tsuboi {et~al.}(2017)Tsuboi, Kitamura, Uehara, Miyawaki, Tsutsumi,
  Miyazaki, \& Miyoshi}]{Tsuboi2017a}
---. 2017, \apj, 842, 94

\bibitem[{Vollmer \& Duschl(2001{\natexlab{a}})}]{Vollmer2001a}
Vollmer, B., \& Duschl, W.~J. 2001{\natexlab{a}}, \aap, 367, 72

\bibitem[{Vollmer \& Duschl(2001{\natexlab{b}})}]{Vollmer2001b}
---. 2001{\natexlab{b}}, \aap, 377, 1016

\bibitem[{Vollmer \& Duschl(2002)}]{Vollmer2002}
---. 2002, \aap, 388, 128

\bibitem[{Wardle \& Yusef-Zadeh(2008)}]{Wardle2008}
Wardle, M., \& Yusef-Zadeh, F. 2008, \apj, 683, L37

\bibitem[{Yusef-Zadeh {et~al.}(2013)Yusef-Zadeh, Royster, Wardle, Arendt,
  Bushouse, Lis, Pound, Roberts, Whitney, \& Wootten}]{Yusef-Zadeh2013}
Yusef-Zadeh, F., Royster, M., Wardle, M., {et~al.} 2013, \apj, 767, L32

\bibitem[{Zhao {et~al.}(2009)Zhao, Morris, Goss, \& An}]{Zhao2009}
Zhao, J.-h., Morris, M.~R., Goss, W.~M., \& An, T. 2009, \apj, 699, 186

\bibitem[{Zubovas {et~al.}(2011)Zubovas, King, \& Nayakshin}]{Zubovas2011}
Zubovas, K., King, a.~R., \& Nayakshin, S. 2011, \mnras, 415, L21

\end{thebibliography}

\end{document}